\documentstyle[12pt]{article}
\def\nabstar#1{\nabla\kern-0.5pt\smash{\raise 4.5pt\hbox{$\ast$}}
               \kern-4.5pt_{#1}}

\def\drvstar#1{\partial\kern-0.5pt\smash{\raise 4.5pt\hbox{$\ast$}}
               \kern-5.0pt_{#1}}

\def\newline{\relax\ifhmode\null\hfil\break\else\nonhmodeerr@\newline\fi}
\def\frac#1#2{{#1\over#2}}
\def\text#1{{\hbox{\rm #1}}}
\def\flushpar{{\par \noindent}}

\newcommand{\beq}{\begin{equation}}
\newcommand{\eeq}{\end{equation}}
\newcommand{\bea}{\begin{eqnarray}}
\newcommand{\eea}{\end{eqnarray}}
\def\Id{ \mbox{1\hspace{-1.2mm}I} }
\def\BE{\begin{equation}}
\def\EE{\end{equation}}
\def\BA{\begin{eqnarray}}
\def\EA{\end{eqnarray}}
\def\BAN{\begin{eqnarray*}}
\def\EAN{\end{eqnarray*}}

\def\nn{\nonumber\\}

\def\tr{\mbox{tr}}
\def\Tr{\mbox{Tr}}
\def\det{\mbox{det}}

\def\gm5{\gamma_5}

\def\CT{{\cal T}}
\def\Dcont{{\cal D}}
\def\anx{{\cal A}(x)}
\def\anxL{{\cal A}_L(x)}
\def\anxw{{\cal A}_W(x)}

\def\anxt{{\cal A}_t(x)}
\def\anxwc{{\cal A}_{wc}(x)}
\input epsf.sty
\newdimen\psfigsize
\def\psfigure#1 #2 #3 #4 #5{
    \begin{figure}[tbh]
      \begin{center}
      \vbox{
        \null\vskip-0.2in\hskip#2
        \epsfxsize=#1
        \epsfbox{#4}
        \vskip -0.3in
        \caption {#5 \label{#3}}
        \vskip 0.0 true in plus 0.3 true in
      }
      \end{center}
   \end{figure}
}
\begin{document}
\thispagestyle{empty}
\begin{flushright}
NTUTH-99-111 \\
November 1999
\end{flushright}
\bigskip\bigskip\bigskip
\vskip 2.5truecm
\begin{center}
{\LARGE {Topological characteristics of
         lattice Dirac operators}}
\end{center}
\vskip 1.0truecm
\centerline{Ting-Wai Chiu}
\vskip5mm
\centerline{Department of Physics, National Taiwan University}
\centerline{Taipei, Taiwan 106, Republic of China.}
\centerline{\it E-mail : twchiu@phys.ntu.edu.tw}
\vskip 2cm
\bigskip \nopagebreak \begin{abstract}
\noindent

We show that even if a lattice Dirac operator satisfies
the conditions consisting of locality, free of species doublings,
correct continuum behavior, $\gm5$-hermiticity and the
Ginsparg-Wilson relation, it does not necessarily have
exact zero modes in nontrivial gauge backgrounds.
This implies that each lattice Dirac operator has its own
topological characteristics which cannot be fixed by these
conditions. The role of topological characteristics in the
axial anomaly is derived explicitly.


\end{abstract}
\vskip 1.5cm

\newpage\setcounter{page}1

\section{ Introduction }

It seems to have been commonly regarded that if a lattice Dirac operator
$ D $ satisfies the conditions : (i) locality, (ii) free of species
doublings, (iii) having correct continuum behavior,
(iv) $\gm5$-hermiticity and (v) the Ginsparg-Wilson relation,
then $ D $ would possess exact zero modes with definite chirality,
obeying the Atiyah-Singer index theorem \cite{AS_thm}
for smooth gauge backgrounds with non-zero integer topological charge.
Explicitly, these conditions are :
\begin{description}
\item[(i)] $ D $ is local. \newline
         ( $ | D_{\alpha\beta}^{ab}(x,y) | \sim \exp( - | x - y | / l ) $
          with $ l \sim a $ and $ l \ll L $ where $ L = N a $ is the
          size of the lattice; or $ D(x,y) = 0 $ for $ |x-y| > z $ with
          $ z \ll L $. )
\item[(ii)] In the free fermion limit, $ D $ is free of species doublings.
             \newline
         ( The free fermion propagator $ D^{-1}(p) $ has only one simple
           pole at the origin $ p = 0 $ in the Brillouin zone. )
\item[(iii)] In the free fermion limit, $ D $ has the correct continuum
             behavior. \newline
         ( In the limit $ a \to 0 $, $ D(p) \sim i \gamma_\mu p_\mu $
           around $ p = 0 $. )
\item[(iv)] $ D $ is $\gm5$-hermitian.
            ( $ D^{\dagger} = \gm5 D \gm5 $. )
\item[(v)] $ D $ breaks the chiral symmetry according to the
             Ginsparg-Wilson relation \cite{gwr}.
           ( $ D \gm5 + \gm5 D = 2 r a  D \gamma_5 D $,
             where $ r $ is a positive real number and
             $ a $ is the lattice spacing. )
\end{description}

However, in this paper, we show that this folklore may not be
justified. We demonstrate that even if a lattice Dirac operator
satisfies all above five conditions {\bf (i)-(v)},
{\it it does not necessarily produce any exact zero modes}
in nontrivial gauge backgrounds. Only for topologically trivial gauge
backgrounds with {\it zero} topological charge, these five conditions
{\bf (i)-(v)} can guarantee that the continuum axial anomaly can be
recovered on the lattice.

The basic motivation of constructing a lattice Dirac operator
satisfying these five conditions {\bf (i)-(v)} is to formulate a
nonperturbatively regularized theory for massless Dirac fermion
interacting with a background gauge field, as a first step towards
the full theory including gauge dynamics.
These five conditions together provide a bypass to the Nielson-Ninomiya
no-go theorem \cite{no-go}, and constitute the necessary conditions
for $ D $ to reproduce the correct physics in the {\it free fermion}
limit as well as in the weak coupling {\it perturbation} regime.
However, they are {\it not} sufficient to
guarantee that any $ D $ satisfying these five conditions must have the
correct index in topologically nontrivial gauge fields.
So far, among all $ D $ satisfying these five conditions {\bf (i)-(v)},
the Neuberger-Dirac operator \cite{hn97:7} is the {\it only}
one which reproduces the correct index ( as well as other important
quantities such as the fermion determinant ratio ). In the next
paragraph, we show how this could be possible in principle, and then
present two explicit counterexamples in Section 3.
Some basic properties of the GW Dirac operator pertinent to our
discussions are collected in the Appendix A.

It has been proved in ref. \cite{twc98:6a} that for any lattice Dirac
operator $ D $ satisfying the $\gm5$-hermiticity {\bf (iv)} and
the GW relation {\bf (v)}, the necessary condition for it to have
nonzero index in the topologically nontrivial gauge background is
\bea
\label{eq:chiu_zenkin}
\det ( \Id - a r D ) = 0.
\eea
In other words, if $ D $ does not satisfy (\ref{eq:chiu_zenkin}),
then the index of $ D $ must be zero. This condition (\ref{eq:chiu_zenkin})
is equivalent to that $ D_c = D ( \Id - r a D )^{-1} $ ( which is
chirally invariant and $\gm5$-hermitian ) has singularities.
Thus, if $ D_c $ is well-defined and does not have any singularities,
then the condition (\ref{eq:chiu_zenkin}) is violated, and the GW Dirac
operator $ D = D_c ( \Id + a r D_c )^{-1} $ must have zero index,
i.e., {\it topologically trivial }.

If one can construct a well-defined $ D_c $ satisfying the four
conditions {\bf (ii)-(iv)} ( note that any $ D_c $ satisfying
conditions {\bf (ii)} and {\bf (iii)} must be nonlocal,
as a consequence of Nielson-Ninomiya theorem \cite{no-go} ),
then the GW Dirac operator $ D = D_c ( \Id + a r D_c )^{-1} $
is topologically trivial and satisfies the four conditions {\bf (ii)-(v)}.
Now if there exists a range of $ r $  such that $ D $ is local,
then $ D $ satisfies all five conditions {\bf (i)-(v)} but has zero
index, hence, a counterexample to the folklore.
We have at least two explicit counterexamples outlined in Section 3.
In general, we can argue that there exists a range of $ r $ such that
$ D $ is local. Our argument is as follows.
First, we note that in the free fermion limit $ D_c $ has a zero
mode at $ p = 0 $ in the Brillouin zone, the condition {\bf (ii)}.
However, unlike any genuine zero modes ( in nontrivial gauge fields )
which are stable against local fluctuations of the gauge fields,
this zero mode can be elevated by turning on a
trivial gauge background with $ A_\mu(x) $ equal to
a tiny constant. Henceforth we shall assume that this trivial
zero mode of $ D_c $ has been elevated in this manner.
Since both $ D $ and $ D_c $ are topologically trivial,
$ D_c $ does not have any zero modes in nontrivial gauge fields.
Thus $ D_c $ does not have any (non-)trivial zero modes.
Then, for fixed lattice spacing $ a $, in the limit
$ r a \gg 1 $,
$ D \simeq (ra)^{-1} \Id + \mbox{ higher order corrections } $,
a highly local operator.
On the other hand, in the limit $ r a \to 0 $,
$ D \to D_c $, which is nonlocal. Thus, between these two extreme
cases, there must exist a range of $ r $ such that $ D $ is local.

Further, since the index of $ D = D_c ( \Id + r a D_c )^{-1} $
is invariant for any $ r $ [ Eq. (\ref{eq:index}) ],
the locality condition {\bf (i)} should {\it not} be
relevant to the question whether $ D $ has the correct index or not.
Only in the case $ D $ has the correct index ( i.e., the sum of the
axial anomaly over all sites is correct ), then the locality of $ D $
comes into the play to ensure that the axial anomaly is also correct
at each site.

So far, we do not know of any additional constraints which
can guarantee that $ D $ has the correct index satisfying
the Atiyah-Singer index theorem.
However, it is obvious that if such a criterion exists, it
cannot be imposed only in the perturbation regime or in the
free fermion limit like conditions {\bf (ii)} and {\bf (iii)},
since it must be pertinent to the topologically nontrivial sectors.

Nevertheless, for any $ D $ satisfying the GW relation {\bf (v)},
its zero modes have definite chirality.
So we can classify any GW Dirac operator $ D $
according to the {\it ratio} of its index ( $ n_{-} - n_{+} $ ) to
the nonzero integer topological charge $ Q $ of the gauge background,
where $ n_{+} $ ( $ n_{-} $ ) denotes the number of zero modes of
positive ( negative ) chirality.
The virtue of such a classification depends on whether
this ratio is {\it robust} with respect to the topological charge $ Q $,
the size of the lattice ( $ L = N a $ ) and
the lattice spacing ( $ a $ ).
It turns out to be the case provided that the number
of sites $ N $ in each direction is sufficiently large
( equivalently, the lattice spacing $ a = L/N $ is sufficiently small )
such that the absolute value of the topological charge
inside any unit cell of volume $ a^4 $ is less than a small number
$ \epsilon_1 $,
\bea
\label{eq:topo_bound}
a^4 | \bar\rho(x) | < \epsilon_1  \hspace{4mm} \forall  x,
\eea
where
\bea
\label{eq:rho_bar}
\bar\rho(x) = \frac{1}{a^4}
              \prod_{\eta=1}^4 \int_{x_\eta - a/2 }^{x_\eta + a/2} d y_\eta
              \ \frac{1}{32 \pi^2}
              \epsilon_{\mu\nu\lambda\sigma}
              \tr ( F_{\mu\nu}(y) F_{\lambda\sigma}(y) ) \ ,
\eea
and $ F $ is the field tensor defined in (\ref{eq:F}).
If one uses $ c[D] $ to denote this ratio, then
\bea
\label{eq:cD}
\mbox{index}(D) =  n_{-} - n_{+} = c[D] \ Q \ ,
\eea
where $ c[D] $ is an {\it integer} constant
provided that (\ref{eq:topo_bound}) is satisfied.
Evidently the value of $ \epsilon_1 $ in
(\ref{eq:topo_bound}) is dependent on $ D $.

In topologically trivial gauge backgrounds
( $ Q = 0 $ ), $ D $ does not have any exact zero modes.
Thus $ c[D] $ cannot be defined by Eq. (\ref{eq:cD}).
Nevertheless, we can define $ c[D] = 1 $ for the trivial sector.
Then the formal definition of $ c[D] $ is
\bea
\label{eq:cD_def}
c[D] = \left\{ \begin{array}{cc}
               ( n_{-} - n_{+} )/Q \ ,  &   Q \ne 0  \\
                     1             \ ,  &   Q = 0
              \end{array}                           \right.
\eea

With gauge backgrounds satisfying the topological
bound (\ref{eq:topo_bound}), we can classify any GW Dirac operator
according to its topological characteristics $ c[D] $ in nontrivial
backgrounds as follows.
If $ c[D] = 1 $, then $ D $ is called {\it topologically proper},
else if $ c[D] = 0 $, then $ D $ is called {\it topologically trivial},
otherwise $ D $ is called {\it topologically improper}.
Evidently, we are only interested in topologically proper $ D $,
since this is a prerequisite for $ D $ to reproduce the
continuum axial anomaly on the lattice.
There are many examples of topologically trivial $ D $, however,
so far, there is only one topologically proper $ D $ which
can be written down explicitly, namely,
the Neuberger-Dirac operator \cite{hn97:7}
\bea
\label{eq:Dh}
D &=& m_0 ( \Id + V ), \hspace{2mm} V = D_w ( D_w^{\dagger} D_w )^{-1/2},
\hspace{2mm} 0 < m_0 < 2 a^{-1} \ , \\
& & D_w = -m_0 + D_W \ ,
\mbox{ $ D_W $ : massless Wilson-Dirac operator } \nonumber
\eea
and its GW generalization
\BAN
D = 2 m_0 ( \Id + V ) [ \Id - V + 2 m_0 r a ( \Id + V ) ]^{-1} \ .
\EAN
Note that we do not have any genuine examples of topologically improper
$ D $ except by setting $ m_0 > 2 a^{-1} $ in (\ref{eq:Dh}) \cite{twc98:10a}.

The outline of this paper is as follows.
In Section 2, we argue that if a lattice Dirac operator does not have
the correct index on {\it any} finites lattices, then
it cannot have the correct index in the continuum limit.
Then we demonstrate that if $ D $ has exact zero modes in a nontrivial
gauge background, then its topological
characteristics can be revealed on a finite lattice,
sometimes even on the smallest lattice of size $ 3^4 $.
We also estimate the topological bound $ \epsilon_1 $ for
the Neuberger-Dirac operator.
In Section 3, we present two examples of lattice Dirac
operators which satisfy the five conditions {\bf (i)-(v)},
but do {\it not} have any exact zero modes in any nontrivial gauge
backgrounds. In Section 4, we derive a general expression
for the axial anomaly of lattice Dirac operator on a
finite lattice, in which the role of its topological
characteristics $ c[D] $ is displayed explicitly.
In Section 5, we conclude and discuss.
In Appendix A, we collect some basic properties of
the GW Dirac operator pertinent to our discussions.
In Appendix B, we show that nonlocal Dirac operators
can have well-defined indices.

\section{ Robustness of the exact zero modes }

It seems to have been commonly believed that a lattice Dirac operator
may possess exact zero modes in the classical continuum limit even if
it does {\it not} have any zero modes in nontrivial gauge
backgrounds on {\it any} finite lattices.
However, our view is that if a lattice Dirac operator
( satisfying {\bf (i)-(v)} ) does not have the correct index
on {\it any} finite lattices, then it will {\it never} have the
correct index in the continuum limit.

First, we set up the notations for our discussions.

Consider a smooth gauge configuration $ \{ A_\mu (x) \} $ with nonzero
integer topological charge $ Q $ in the 4-dimensional Euclidean spacetime
which is temporarily truncated to a hypercube of size $ L^4 $
with periodic boundary conditions, i.e., a 4-d flat torus, $ T^4 $.
( The infinite volume limit $ L \to \infty $
is presumably to be taken at the end of all calculations. )

The covariant differential operator $ \Dcont_\mu $ acting on the
matter fields is defined as
$$
\Dcont_\mu =  \partial_\mu + i A_\mu \ .
$$
The field tensor $ F_{\mu\nu} $ is defined by the commutator,
\bea
\label{eq:F}
[ \Dcont_\mu, \Dcont_\nu ] = i F_{\mu\nu} \ .
\eea
The topological charge density of the gauge background is
\bea
\label{eq:rho}
\rho (x) = \frac{1}{32 \pi^2} \epsilon_{\mu\nu\lambda\sigma}
                              \tr ( F_{\mu\nu}(x) F_{\lambda\sigma}(x) )
\eea
where $ \tr $ denotes the trace over the gauge group space.
Then the topological charge
\bea
\label{eq:Q_top}
Q  = \int_{T^4} d^4 x \ \rho(x)
   = \int_{T^4} d^4 x \ \frac{1}{32 \pi^2} \epsilon_{\mu\nu\lambda\sigma}
                        \tr ( F_{\mu\nu}(x) F_{\lambda\sigma}(x) ) \
\eea
is an integer.

The massless Dirac operator in continuum is
$$
\Dcont = \gamma_\mu \Dcont_\mu = \gamma_\mu ( \partial_\mu + i A_\mu ) \ ,
$$
which is chirally invariant
( $ \Dcont \gamma_5 + \gamma_5 \Dcont = 0 $ )
and antihermitian ( $ \Dcont^{\dagger} = - \Dcont $ ).
The Dirac matrices are defined by
\bea
\label{eq:gamma_mu}
\gamma_\mu = \left( \begin{array}{cc}
                            0                &  \sigma_\mu    \\
                    \sigma_\mu^{\dagger}     &       0
                    \end{array}  \right)
\eea
where
\beq
\label{eq:sigma_mu}
 \sigma_{\mu} \sigma_{\nu}^{\dagger} + \sigma_{\nu} \sigma_{\mu}^{\dagger}
= 2 \delta_{\mu\nu} \ .
\eeq
Explicitly, we choose $ \sigma_1, \sigma_2, \sigma_3 $ to be the
Pauli matrices, and $ \sigma_4 = i \Id $ where $ \Id $ is the
$ 2 \times 2 $ unit matrix. Then the chirality operator is
\beq
\label{eq:gamma5}
\gamma_5 = \gamma_1 \gamma_2 \gamma_3 \gamma_4
         = \left( \begin{array}{cc}
                        \Id    &    0    \\
                         0     &  -\Id
                  \end{array}  \right) \ .
\eeq

In a smooth background gauge field with nonzero integer topological
charge $ Q $, $ \Dcont $ has zero eigenvalues and the corresponding
eigenfunctions are chiral. The Atiyah-Singer index theorem \cite{AS_thm}
asserts that the difference of the number of left-handed and right-handed
zero modes is equal to the topological charge of the gauge configuration,
\bea
\label{eq:index_thm}
n_{-} - n_{+} = Q \ .
\eea

The lattice regularization of the gauge background
amounts to discretize the 4-d torus into a hypercubical
lattice with $ N $ sites in each direction ( i.e., $ L = N a $ ),
and then transcribe the gauge
fields  $ \{ A_\mu (x) \} $ to the link variables
$ \{ U_\mu (x) \} $, where $ U_\mu(x) $ is the link variable
defined on the link $ (x,x + a \hat\mu ) $.
We choose the lattice spacing $ a $ sufficiently small
( i.e., $ N $ is sufficiently large for fixed $ L $. )
such that the variations of $ \{ A_\mu (x) \} $
can be tracked by the link variables $ \{ U_\mu (x) \} $.
For any lattice Dirac operator $ D $
( satisfying the five conditions {\bf (i)-(v)} ), our primary
concern is whether $ D $ can produce exact zero modes satisfying the
Atiyah-Singer index theorem (\ref{eq:index_thm}).
If the index of $ D $ always agrees with the topological
charge $ Q $ for any gauge background
satisfying the bound (\ref{eq:topo_bound}),
then $ D $ is called topologically proper.
It is evident that the value of $ \epsilon_1 $ depends on $ D $.
A rough estimate of $ \epsilon_1 $ can be obtained by the following
simple prescription.

Consider the following gauge configuration with
constant field tensor and topological charge $ Q = 1 $,
\BAN
A_1^{c} (x) &=& - \frac{ 2 \pi x_2 }{L^2} \  t^{\alpha}, \\
A_2^{c} (x) &=& 0, \\
A_3^{c} (x) &=& - \frac{ 2 \pi x_4 }{L^2} \  t^{\alpha}, \\
A_4^{c} (x) &=& 0
\EAN
where $ t^{\alpha} $ is any one of the generators of the gauge group
with normalization $ \tr ( t^{\alpha} t^{\beta} ) = \delta^{\alpha\beta} $.
The nonzero components of the field tensor are
\bea
F_{12}(x) = F_{34}(x) = \frac{ 2 \pi }{L^2} \ t^{\alpha} \equiv F \ .
\eea
Then the corresponding link variables on the lattice
( $ x_\mu = 0, \cdots, (N-1)a $ ) are :
\bea
\label{eq:U1a}
U_1(x) &=&
    \exp \left[ \ -\text{i} \frac{2\pi a x_2}{L^2} \ t^{\alpha} \ \right] \\
\label{eq:U2a}
U_2(x) &=& \exp \left[ \ \text{i}\delta_{x_2,(N-1)a} \
    \frac{ 2 \pi x_1 }{L} \ t^{\alpha} \ \right] \\
\label{eq:U3a}
U_3(x) &=&
   \exp \left[ \ -\text{i} \frac{2\pi a x_4}{L^2} \ t^{\alpha} \ \right] \\
\label{eq:U4a}
U_4(x) &=& \exp \left[ \ \text{i} \delta_{x_4,(N-1)a} \
                \frac{ 2 \pi x_3 }{L} \ t^{\alpha} \ \right] \ .
\eea
Using the notation
\bea
\label{eq:Umn}
U_{\mu\nu}(x) = U_\mu (x) U_\nu (x + a \hat\mu )
                U_{\mu}^{\dagger} ( x + a \hat\nu )
                U_{\nu}^{\dagger} ( x )
\eea
to stand for the path-ordered product of link variables
around the plaquette $ p = ( x, \hat\mu, \hat\nu ) $, we obtain
\bea
U_{12}(x) = U_{34}(x) =
\exp \left[ \ \text{i} \frac{2\pi a^2 }{L^2} \ t^{\alpha} \ \right]
= \exp \left[ \ \text{i} a^2 F \ \right]
\eea
and $ U_{13}(x) = U_{14} (x) = U_{23}(x) = U_{24}(x) = \Id $.

Then we test whether $ D $ has exact zero modes on finite lattices by
successively increasing the number of sites $ N $
in the sequence $ N = 3, 4, 5, \cdots $.

If $ D $ is topologically proper, then the exact zero mode
would emerge after $ N > N_1 $ for a certain finite integer $ N_1 $,
( i.e., for the strength of the nonzero field tensor inside
each plaquette, $ a^2 | F_{12} | = a^2 | F_{34} | =  2 \pi / N^2 $,
is less than $ 2 \pi / N_1^2  $ ).
This yields a rough estimate of
$ \epsilon_1 \simeq 1/(N_1+1)^4 $ for the topological bound
(\ref{eq:topo_bound}) such that $ c[D] $ is an integer constant.
A more optimal estimate of $ \epsilon_1 $ can be obtained by
introducing local fluctuations which change the topological charge
density locally but maintain the total topological charge fixed.

On the other hand, if $ D $ is topologically trivial, then $ D $
does not have any exact zero modes for any {\it finite} lattice
spacing $ a = L/N $ ( as $ N $ is increased successively ).
Suppose that in the limit $ a = 0 $ ( $ N = \infty $ ),
the exact zero mode emerges, however, in our view,
$ D $ still does {\it not} have the correct index in the continuum limit.
Our argument is as follows.

\flushpar
[ Proof ] :

First, let the numbers of exact zero modes of $ D $ be
$ k_{+} $ and $ k_{-} $ in a nontrivial gauge background with
topological charge $ Q $ on a finite lattice,
and they do not yield the correct index.
( i.e., $ \mbox{index}(D) = k_{-} - k_{+} \ne Q $. )
Suppose that the index of $ D $ remains the same for any
finite lattice spacing $ a = L/N $ as $ N $ is increased successively,
except at $ a = 0 $ where the correct numbers of exact zero modes
( $ n_{\pm} $ ) emerge and the index becomes $ n_{-} - n_{+} = Q $.
This implies that the index of $ D $ must undergo a {\it discontinuous}
transition from the integer $ k_{-} - k_{+} $ to another integer
$ n_{-} - n_{+} $ at $ a = 0 $, and correspondingly, $ D $ becomes $ D_1 $.
Thus the $ \mbox{index}(D_1) = n_{-} - n_{+} $ does not
constitute a limiting point of the $ \mbox{index}(D) = k_{-} - k_{+} $,
because a discontinuity has occurred.
Since the index of $ D_1 $ is not equal to the index of $ D $,
they are in two different topological phases,
or in other words, not in the same universality class.
It follows that $ D_1 $ cannot be the continuum limit of $ D $.
This completes the proof.

In short, {\it if $ D $ does not have the correct index
at finite lattice spacing, then $ D $ cannot have the
correct index in the continuum limit.}

Therefore we conclude that the index of a lattice Dirac operator
( satisfying the conditions {\bf (i)-(v)} )
is invariant for any nontrivial gauge backgrounds with the
same topological charge provided that the topological
bound (\ref{eq:topo_bound}) is satisfied.

Next we demonstrate that if a lattice Dirac operator is topologically
proper, then it can produce the correct index on a {\it finite} lattice,
sometimes even on a very small lattice.
For example, on the $ 3^4 $ lattice, the Neuberger-Dirac operator
(\ref{eq:Dh}) with $ m_0 = 1 $ has one exact zero mode in a gauge
background with $ Q = 1 $. We also estimate the value of its
$ \epsilon_1 $ in the topological bound (\ref{eq:topo_bound}).

For simplicity, we consider the following $ U(1) $
background gauge fields on the 4-d flat torus
( $ x_\mu \in [ 0, L ], \ \mu = 1, \cdots, 4 $ ) :
\bea
\label{eq:A1}
a A_1(x) &=&
                        - \frac{ 2 \pi q_1 a x_2 }{ L^2 }
     +  A_1^{(0)} \sin \left( \frac{ 2 \pi n_2 }{L} x_2 \right) \\
\label{eq:A2}
a A_2(x) &=&
        A_2^{(0)} \sin \left( \frac{ 2 \pi n_1 }{L} x_1 \right) \\
\label{eq:A3}
a A_3(x) &=&
               - \frac{ 2 \pi q_2 a x_4 }{ L^2 }
     +  A_3^{(0)} \sin \left( \frac{ 2 \pi n_4 }{L} x_4 \right) \\
\label{eq:A4}
a A_4(x) &=&
        A_4^{(0)} \sin \left( \frac{ 2 \pi n_3 }{L} x_3 \right)
\eea
where $ q_1 $ and $ q_2 $ are integers.
The global part of the gauge background is characterized by the
topological charge
$$
Q = \frac{1}{32\pi^2} \int d^4 x \ \epsilon_{\mu\nu\lambda\sigma} \
     F_{\mu\nu} (x) F_{\lambda\sigma} (x) =  q_1 \ q_2 \ .
$$
The local parts are chosen to be sinusoidal
fluctuations with amplitudes $ A_1^{(0)} $, $ A_2^{(0)} $, $ A_3^{(0)} $
and $ A_4^{(0)} $, and
frequencies $ \frac{ 2 \pi n_2 }{L} $, $ \frac{ 2 \pi n_1 }{L} $,
$ \frac{ 2 \pi n_4 }{L} $ and $ \frac{ 2 \pi n_3 }{L} $ where
$ n_1 $, $ n_2 $, $ n_3 $ and $ n_4 $  are integers.
The discontinuity of $ A_1(x) $ ( $ A_3(x) $ ) at $ x_2 = L $
( $ x_4 = L $ ) due to the global part only amounts to a gauge
transformation. The components of the field tensor
are continuous on the flat torus.
To transcribe the background gauge field to link variables on the lattice,
we take the lattice sites at $ x_\mu = 0, a, ..., ( N - 1 ) a $,
where $ a $ is the lattice spacing and $ L = N a $ is the lattice size.
Then the link variables are
\bea
\label{eq:U1}
U_1(x) &=& \exp \left[ \ \text{i} a A_1(x) \ \right] \\
\label{eq:U2}
U_2(x) &=& \exp \left[ \ \text{i} a A_2(x)
 + \text{i} \delta_{x_2,(N - 1)a} \frac{ 2 \pi q_1 x_1 }{L} \ \right] \\
\label{eq:U3}
U_3(x) &=& \exp \left[ \ \text{i} a A_3(x) \ \right] \\
\label{eq:U4}
U_4(x) &=& \exp \left[ \ \text{i} a A_4(x)
 + \text{i} \delta_{x_4,(N - 1)a} \frac{ 2 \pi q_2 x_3 }{L} \ \right]
\eea

First, we consider the simplest nontrivial gauge configuration with
constant field tensor at each plaquette, i.e.,
\beq
a^2 F_{\mu\nu} (x) = - a^2 F_{\nu\mu} (x) = \left\{ \begin{array}{ll}
         2 \pi q_1 / N^2,  &  \{ \mu, \nu \} = \{ 1, 2 \} \ ;\\
         2 \pi q_2 / N^2,  &  \{ \mu, \nu \} = \{ 3, 4 \} \ ; \\
                0,         &  \mbox{ otherwise. } \\
                              \end{array}
                              \right.
\label{eq:Fmn}
\eeq
This amounts to setting the local sinusoidal fluctuations
to zero (  $ A_i^{(0)} = 0 $, $ i = 1, \cdots, 4 $ ) in
Eqs. (\ref{eq:A1})-(\ref{eq:A4}).

In Table 1, we list the number of exact zero modes of each chirality
versus the topological charge of the simplest nontrivial gauge
background, for the Neuberger-Dirac operator on the $ 3^4 $ lattice.
The Atiyah-Singer index theorem (\ref{eq:index_thm}) is satisfied
exactly for $ Q = 1 $, but not for $ Q = 2 $. This yields
a rough estimate of $ \epsilon_1 $ in (\ref{eq:topo_bound}) :
$ 1/3^4 < \epsilon_1 < 2/3^4 $.
It is obvious that the Neuberger-Dirac operator
is {\it nonlocal} on such a small lattice. This demonstrates
that even a {\it nonlocal Dirac operator can have exact zero modes
satisfying the Atiyah-Singer index theorem.}
However, the axial anomaly of a nonlocal lattice Dirac operator
must disagree with the topological charge density of the gauge
background.

Besides the zero modes, the $+2$ eigenmodes of the Neuberger-Dirac
operator are also chiral \cite{twc98:4}, and the numbers of both
chiralities are listed in the last two columns in Table 1.
It has been shown that for any $ D $ satisfying
conditions {\bf (iv) and (v)}, every zero mode of $ D $ must be
accompanied by a $ 1/r $ real eigenmode of opposite chirality,
and they satisfy the chirality sum-rule \cite{twc98:4},
\bea
\label{eq:chi_sum}
n_{-} + n_2^{-} = n_{+} + n_2^{+} \ .
\eea
where $ r=1/2 $ for the Neuberger-Dirac operator.
In Table 1, it is evident that the zero modes together with
the $+2$ chiral modes satisfy the chirality sum-rule (\ref{eq:chi_sum}).

{\footnotesize
\begin{table}
\begin{center}
\begin{tabular}{|c|c|c|c|c|c|c|c|}
\hline
$ q_1 $ & $ q_2 $ & $ Q = q_1 q_2 $ & $ n_{+} $ & $ n_{-} $ &
$ n_2^{+} $ & $ n_2^{-} $ & $ c[D] $ \\
\hline
\hline
  -1  &    -1   &     1   &     0   &    1   &  1  &  0  &  1  \\
\hline
  -1  &     1   &    -1   &     1   &    0   &  0  &  1  &  1  \\
\hline
   1  &     1   &     1   &     0   &    1   &  1  &  0  &  1  \\
\hline
\hline
  -2  &     1   &    -2   &     1   &    0   &  0  &  1  &  1/2 \\
\hline
   2  &    -1   &    -2   &     1   &    0   &  0  &  1  &  1/2 \\
\hline
  -2  &    -1   &    -2   &     0   &    1   &  1  &  0  &  1/2 \\
\hline
   2  &     1   &     2   &     0   &    1   &  1  &  0  &  1/2 \\
\hline
\end{tabular}
\end{center}
\caption{The exact zero modes versus the topological charge,
for the Neuberger-Dirac operator (\ref{eq:Dh}) with $ m_0 = 1 $,
on the $ 3^4 $ lattice.
The Atiyah-Singer index theorem $ Q = n_{-} - n_{+} $ is satisfied
only for $ | Q | = 1 $. The chirality sum rule (\ref{eq:chi_sum})
is satisfied in all cases.  The last column is
the topological characteristics, $ c[D] = ( n_{-} - n_{+} )/Q $.
The fractional value of $ c[D] $ indicates that topological bound
is violated.}
\label{table:1}
\end{table}
}

Note that the exact zero modes ( as well as the $+2$ chiral modes )
are very stable under local fluctuations of the gauge
background provided that the topological bound (\ref{eq:topo_bound}) is
satisfied. ( Otherwise, they are not regarded as genuine zero modes
of a lattice Dirac operator in a nontrivial gauge background. )
To demonstrate the robustness of the zero modes, as well as to obtain
a better estimate of $ \epsilon_1 $, we turn on the local sinusoidal
fluctuations in (\ref{eq:A1}).
With $ Q = 1 $ on the $ 3^4 $ lattice, we increase the amplitude
$ A_1^{(0)} $ gradually until the exact zero mode disappears.
At this point, the average of $ a^4 | \bar\rho(x) | $
(\ref{eq:rho_bar}) over all sites gives a conservative estimate
of $ \epsilon_1 $.
In Table 2, the amplitude $ A_1^{(0)} $ is listed in the first column.
The frequency of the sinusoidal fluctuation is $ 2 \pi / L $.
The maximum and the average of $ a^4 | \bar\rho(x) | $
are listed in the second and third columns respectively.
We note that for $ 0 \le A_1{(0)} < 1.022 $, the index is
$ n_{-} - n_{+} = 1 $, equal to the topological charge $ Q = 1 $.
However, as soon as $ A_1^{(0)} $ reaches $ 1.022 $, the
exact zero mode disappears and the index becomes zero.
As we further increase $ A_1^{(0)} $ up to $ 1.090 $,
$ D $ enters into another topological phase with index equal to $ -1 $.
Thus, at $ A_1^{(0)} = 1.022 $, the average of $ a^4 | \bar\rho(x) | $
over all sites yields a conservative estimate of
$ \epsilon_1 < 0.025 $. To be secure, we take $ \epsilon_1 = 0.02 $.
For various gauge configurations on larger lattices, we have checked
that the index of the Neuberger-Dirac operator with $ m_0 = 1 $ is
always equal to the topological charge of the background provided that
the topological bound
\bea
\label{eq:bound_1_Dh}
a^4 | \bar\rho(x) | < 0.02 \hspace{4mm} \forall \ x
\eea
is satisfied.

{\footnotesize
\begin{table}
\begin{center}
\begin{tabular}{|c|c|c|c|c|c|c|c|}
\hline
$ A_1^{(0)} $ & $ \mbox{max}( a^4 | \bar\rho(x) |) $ & $ < a^4 \bar\rho(x) > $
& $ n_{+} $ & $ n_{-} $ & $ n_2^{+} $ & $ n_2^{-} $ & $ c[D] $\\
\hline
\hline
  0.000  &  0.0123   &  0.0123   &     0   &    1   &  1  &  0  &  1 \\
\hline
  0.400  &  0.0185   &  0.0123   &     0   &    1   &  1  &  0  &  1 \\
\hline
  0.600  &  0.0215   &  0.0164   &     0   &    1   &  1  &  0  &  1 \\
\hline
  0.800  &  0.0246   &  0.0205   &     0   &    1   &  1  &  0  &  1 \\
\hline
  1.000  &  0.0277   &  0.0245   &     0   &    1   &  1  &  0  &  1 \\
\hline
  1.020  &  0.0279   &  0.0249   &     0   &    1   &  1  &  0  &  1 \\
\hline
\hline
  1.022  &  0.0280   &  0.0250   &     0   &    0   &  0  &  0  &  0 \\
\hline
  1.080  &  0.0289   &  0.0262   &     0   &    0   &  0  &  0  &  0 \\
\hline
  1.085  &  0.0289   &  0.0263   &     0   &    0   &  0  &  0  &  0 \\
\hline
\hline
  1.090  &  0.0290   &  0.0264   &     1   &    0   &  0  &  1  &  -1 \\
\hline
  1.100  &  0.0292   &  0.0266   &     1   &    0   &  0  &  1  &  -1 \\
\hline
\end{tabular}
\end{center}
\caption{The exact zero modes versus the amplitude of the local
sinusoidal fluctuations, $ A_1^{(0)} $ in (\ref{eq:A1}),
for the Neuberger-Dirac operator on the $ 3^4 $ lattice.
The topological charge is $ Q = 1 $.
The Atiyah-Singer index theorem $ Q = n_{-} - n_{+} $ is satisfied
for $ A_1^{(0)} < 1.022 $, or equivalently,
$ a^4 | \bar\rho(x) | < 0.028 $. The last column is
the topological characteristics, $ c[D] = ( n_{-} - n_{+} )/Q $. }
\label{table:2}
\end{table}
}

We note that the topological bound (\ref{eq:topo_bound})
can be satisfied if the link configuration $ \{ U_\mu(x) \} $
fulfils
\bea
\label{eq:U_bound_1}
| \mbox{Re} \ \tr(\Id - U_{\mu\nu}(x) )|
< \frac{ 2 \pi^2 \epsilon_1 }{3} \hspace{2mm}
\mbox{ for all plaquettes, }
\eea
where $ U_{\mu\nu}(x) $ is the path-ordered product of link variables
around the plaquette $ p=( x, \hat\mu, \hat\nu ) $, as defined in
(\ref{eq:Umn}).
However, the relevant parameter for the topological phase
transition of $ D $ is the topological charge density
$ a^4 | \bar\rho(x) | $ rather than the
plaquette action $ |\mbox{Re} \ \tr(\Id - U_{\mu\nu}(x) ) | $,
since a topologically proper $ D $ always has $ c[D] = 1 $
for any gauge configuration satisfying (\ref{eq:topo_bound})
even if it violates (\ref{eq:U_bound_1}) for some plaquettes.

Finally, we note that the topological bound (\ref{eq:topo_bound})
or (\ref{eq:U_bound_1}) does not guarantee that $ D $ is local,
but only that $ c[D] $ is an integer constant.
In general, if a lattice Dirac operator is local
in the free fermion limit, then it would maintain its localness
in a gauge background with sufficiently small field strength at
each plaquette, i.e.,
\bea
\label{eq:U_bound_2}
|| \Id - U_{\mu\nu} (x) || < \epsilon \hspace{2mm}
\mbox{ for all plaquettes, }
\eea
where the matrix norm $ || M || $ can be defined to be
the square root of the maximum eigenvalue of $ M^{\dagger} M $,
i.e., $ || M || = \sqrt{ \lambda_{max} (  M^{\dagger} M ) } $.
Evidently the value of $ \epsilon $ is dependent on $ D $.

The locality bound (\ref{eq:U_bound_2}) implies that the topological
charge inside any unit cell satisfies
\bea
\label{eq:rho_bound_2}
a^4 | \bar\rho(x) | < \epsilon_2 \
\hspace{4mm} \forall \ x \ , \\
\epsilon_2 =  \frac{3}{4\pi^2} \tr( \Id ) \epsilon^2
\eea
where $ \tr(\Id) = N_c $ for link variables in the fundamental
representation of the gauge group $ SU(N_c) $.
For the Neuberger-Dirac operator with $ m_0 = 1 $,
$ \epsilon = 1/6(2+\sqrt{2}) \simeq 0.0488 $ \cite{hn99:11a,ml98:8a},
which gives $ \epsilon_2 \simeq 5.43 \times 10^{-4} $ for $ N_c = 3 $.
This shows that $ \epsilon_2 $ is much less than
$ \epsilon_1 \simeq 0.02 $ in the topological bound.
Evidently, for any topologically proper $ D $,
the inequality $ \epsilon_2 < \epsilon_1 $ must hold,
since the topological bound only requires that the index
( i.e., the sum of the axial anomaly over all sites ) is equal
to the topological charge, while the locality bound further
guarantees that the axial anomaly $ \anxL $ at each site is in good
agreement with the topological charge density $ \rho(x) $ of the gauge
background.

We summarize the main theme of this section as follows.
Consider a fixed gauge background with nonzero integer
topological charge $ Q $ on the 4-torus of size $ L^4 $.
Then we impose a lattice on the 4-torus with lattice spacing $ a $,
and the number of sites in each direction, $ N $ ( i.e., $ L = N a $ ).
Given any lattice Dirac operator $ D $,
we can test whether the index of $ D $
agrees with $ Q $ by successively increasing
$ N $ ( equivalently, decreasing lattice spacing $ a = L/N $ )
in the sequence $ N = 3,4, \cdots $. If $ D $ is topologically proper,
then there exists a finite integer $ N_1 $ such that
for $ N > N_1 $ ( $ a < L/N_1 $ ), the index of $ D $ is equal to $ Q $.
However, when $ N $ is only slightly larger than $ N_1 $
( equivalently, $ a $ is slightly less than $ L/N_1 $ ),
$ D $ may not be local yet.
In general, if $ D $ is local in the free fermion limit, then
there exists another integer $ N_2 > N_1 $ such that
for $ N > N_2 $ ( $ a < L/N_2 $ ), $ D $ is local.
However, if $ D $ is topologically trivial, then $ D $ does not have
any zero modes for any finite $ N $, even though it has become local
for $ N > N_2 $. Whether $ D $ has any zero modes at
$ N = \infty $ ( $ a = 0 $ ) really does not matter at all,
since in case it has, then it cannot be the continuum limit of $ D $.

So far, among the lattice Dirac operators which satisfy the five
conditions {\bf (i)-(v)}, the Neuberger-Dirac operator is the {\it only}
one which is {\it topologically proper} for gauge configurations
satisfying the topological bound : $ a^4 | \bar\rho(x) | < 0.02 $.
In the next section, we present two explicit examples which satisfy
these five conditions {\bf (i)-(v)}, but do {\it not} have any exact
zero modes in nontrivial gauge backgrounds.

\section{ Topologically trivial lattice Dirac operators }

It has been proved in ref. \cite{twc98:6a} that for any chirally
symmetric and $\gm5$-hermitian Dirac operator $D_c $, if it
is well-defined ( without any singularities ), then the GW Dirac
operator $ D = D_c ( \Id + r a D_c )^{-1} $ has zero index in any
nontrivial gauge background.

Therefore, if one can construct a well-defined $ D_c $ satisfying
the three conditions {\bf (ii)-(iv)} ( note that any $ D_c $ satisfying
{\bf (ii)-(iii)} must be nonlocal, as a consequence of no-go theorem ),
then the GW Dirac operator $ D = D_c ( \Id + a r D_c )^{-1} $
is topologically trivial and satisfies all five conditions {\bf (i)-(v)}.
The locality of $ D $ in the free fermion is ensured
by choosing $ r $ in the proper range. Then it would be
local in a gauge background which satisfies the locality bound
(\ref{eq:U_bound_2}). In this section, we present two examples from
recent studies \cite{twc99:8, twc99:12a}.

In ref. \cite{twc99:8}, a well-defined $ D_c $ satisfying the three
conditions {\bf (ii)-(iv)} has been constructed,
\bea
D_c
\label{eq:Dc1}
\equiv \left[ \begin{array}{cc}
                     0          &   -D_L^{\dagger}  \\
                     D_L        &   0
              \end{array}                           \right] \ ,
\eea
where
\bea
\label{eq:D_L}
D_L &=& ( \sigma \cdot t )^{-1}
  \left[ w - \sqrt{ w^2 } \sqrt{ 1 + w^{-1} \ t^2 \ w^{-1} } \ \right] \ , \\
\label{eq:sigma_t}
\sigma \cdot t &=& \sum_\mu \sigma_\mu t_\mu  \ ,  \\
\label{eq:tmu}
t_\mu (x,y) &=& \frac{1}{2} \ [   U_{\mu}(x) \delta_{x+\hat\mu,y}
                       - U_{\mu}^{\dagger}(y) \delta_{x-\hat\mu,y} ] \ , \\
\label{eq:t2}
t^2 &=& - ( \sigma \cdot t ) ( \sigma^{\dagger} \cdot t ) \ ,
\eea
\BAN
\sigma_\mu \sigma_\nu^{\dagger} + \sigma_{\nu} \sigma_\mu^{\dagger} =
2 \delta_{\mu \nu} \ , \\
\gamma_\mu = \left( \begin{array}{cc}
                            0                &  \sigma_\mu    \\
                    \sigma_\mu^{\dagger}     &       0
                    \end{array}  \right) \ ,
\EAN
and
\bea
\label{eq:wxy}
w(x,y) = \delta_{x,y} - \frac{1}{2} \sum_\mu \left[ 2 \delta_{x,y}
                       - U_{\mu}(x) \delta_{x+\hat\mu,y}
                       - U_{\mu}^{\dagger}(y) \delta_{x-\hat\mu,y} \right] \ .
\eea

It is evident that $ D_c $ (\ref{eq:Dc1}) is $\gm5$-hermitian
since it is chirally symmetric and antihermitan.
It has been shown \cite{twc99:8} that
in the free fermion limit, $ D_c $ (\ref{eq:Dc1})
is free of species doublings at finite lattice spacing and agrees with
$ \gamma_\mu \partial_\mu $ in the classical continuum limit.
So, $ D_c $ (\ref{eq:Dc1}) satisfies the three conditions {\bf (ii)-(iv)}.

Now we examine whether $ D_c $ ( $ D_L $ ) is well-defined.
Since the naive lattice Dirac fermion operator $ \gamma_\mu t_\mu $
does not have any exact zero modes in nontrivial gauge backgrounds,
its left-handed fermion propagator $ ( \sigma \cdot t )^{-1} $ is
well-defined. Therefore, the first factor of $ D_L $ (\ref{eq:D_L})
is well defined. Next we examine the second factor of
$ D_L $ (\ref{eq:D_L}).
We see that the second factor is well-defined except for the gauge
configurations which give $ \det( w ) = 0 $. However, these
exceptional configurations have zero measure, thus they would
never be encountered in any practical calculations.
Therefore we have confirmed that $ D_c $ (\ref{eq:Dc1}) is well
defined and satisfies the three conditions {\bf (ii)-(iv)}.
Then the GW Dirac operator $ D = D_c ( \Id + a r D_c )^{-1} $
satisfies the five conditions {\bf (i)-(v)}, where the locality of $ D $
is ensured by choosing $ r $ in the proper
range. Since $ D_c $ is well defined, $ D = D_c ( \Id + a r D_c )^{-1} $
is topologically trivial, according to the theorem proved
in ref. \cite{twc98:6a}. Explicitly,
\bea
\label{eq:Dtwc}
 D = \left[ \begin{array}{cc}
 r ( b b^{\dagger} + r^2 )^{-1}  &  - b ( b^{\dagger} b + r^2 )^{-1}   \\
 b^{\dagger} ( b b^{\dagger} + r^2 )^{-1}  &  r ( b^{\dagger} b + r^2 )^{-1}
            \end{array}      \right]
\eea
where
\bea
\label{eq:Bwt}
b &=& \frac{1}{2} \left( w - \sqrt{w^2} \sqrt{\Id + w^{-1} \ t^2 \ w^{-1} } \
                  \right)^{-1} \ ( \sigma \cdot t )
\eea

Another example of well-defined $ D_c $ which satisfies the three
conditions {\bf (ii)-(iv)} is constructed in ref. \cite{twc99:12a},
\bea
\label{eq:Dwc}
D_c = \gamma_\mu t_\mu - W \ ( \gamma_\mu t_\mu )^{-1} \ W  \ .
\eea
where $ t_\mu $ is defined in (\ref{eq:tmu}) and $ W $ is
the Wilson term \cite{wilson75}
\bea
\label{eq:wilson}
W(x,y) =  \frac{1}{2} \sum_\mu \left[ 2 \delta_{x,y}
                     - U_{\mu}(x) \delta_{x+\hat\mu,y}
                     - U_{\mu}^{\dagger}(y) \delta_{x-\hat\mu,y} \right] \ .
\eea
The $ D_c $ in (\ref{eq:Dwc}) is $\gm5$-hermitian, since it is
chirally symmetric and antihermitian.
It is nonlocal due to the factor $ ( \gamma_\mu t_\mu )^{-1} $ in
the second term. But it is well-defined since the naive fermion
operator $ \gamma_\mu t_\mu $ does not have any exact zero modes in
nontrivial gauge backgrounds. The free fermion propagator of
$ D_c $ (\ref{eq:Dwc}) in momentum space is
\bea
\label{eq:Dci}
D_c^{-1} (p) = ( \gamma_\mu t_\mu )^{-1} \frac{ t^2 }{ w^2 + t^2 }
\eea
where $ t_\mu = i a^{-1} \sin( p_\mu a ) $,
      $ t^2 = a^{-2} \sum_{\mu} \sin^2 ( p_\mu a ) $
and   $ w = 2 a^{-1}  \sum_{\mu} \sin^2( p_\mu a /2 ) $.
The doubled modes are decoupled completely due to the
vanishing of the factor $ t^2/( w^2 + t^2 ) $ at the $ 2^d - 1 $
corners of the Brillouin zone. So the fermion propagator $ D_c^{-1}(p) $
is free of species doublings. In the limit $ a \to 0 $, Eq. (\ref{eq:Dci})
gives $ D_c (p) = i \gamma_\mu p_\mu $ around $ p = 0 $.
Now we have confirmed that $ D_c $ (\ref{eq:Dwc}) is well-defined,
and it satisfies the conditions {\bf (ii)-(iv)}.

Then we substitute (\ref{eq:Dwc}) into the formula
$ D = D_c ( \Id + a r D_c )^{-1} $ to obtain a GW Dirac operator
\bea
\label{eq:Dw_GW}
D = \left[ \begin{array}{cc}
           r \ C^{\dagger} C ( \Id  + r^2 C^{\dagger} C )^{-1}  &
           - C^{\dagger} ( \Id + r^2 C C^{\dagger} )^{-1}      \\
           C ( \Id + r^2 C^{\dagger} C )^{-1}  &
           r \ C C^{\dagger} ( \Id + r^2 C C^{\dagger} )^{-1}
                                             \end{array}      \right]
\eea
where
\bea
\label{eq:C}
C = ( \sigma^{\dagger}_\mu t_\mu ) - W ( \sigma_\mu t_\mu )^{-1} W \ .
\eea
Then $ D $ (\ref{eq:Dw_GW}) satisfies all five conditions {\bf (i)-(v)},
where the locality of $ D $ is ensured by choosing $ r $ in the proper
range. Since $ D_c $ is antihermitian and well-defined, the index of
$ D = D_c ( \Id + r a D_c ) $ must be zero \cite{twc98:6a}.

We have explicitly checked that the GW Dirac operators in
(\ref{eq:Dtwc}) and (\ref{eq:Dw_GW}) do not have any exact zero
modes in a constant nontrivial gauge background with $ Q = 1 $,
[ Eqs. (\ref{eq:U1a})-(\ref{eq:U4a}) ], on different lattices with
sizes ranging from $ N = 3 $ to $ N = 6 $.

We note that even though these two GW Dirac operators (\ref{eq:Dtwc})
and (\ref{eq:Dw_GW}) do not have any exact zero modes in nontrivial gauge
backgrounds, the five conditions {\bf (i)-(v)} do guarantee that
they can reproduce the correct axial anomaly in {\it trivial} gauge
backgrounds. This will be demonstrated explicitly in the next section.

\section{ The axial anomaly and topological characteristics }

In this section, we derive a general expression
for the axial anomaly of lattice Dirac operator $ D $ on a
finite lattice, in which the role of the topological
characteristics $ c[D] $ is displayed explicitly.

\subsection{ In continuum }

First, we discuss the axial anomaly of the massless Dirac operator
in a gauge background in the continuum, i.e., the 4-d flat torus ( $ T^4 $ )
with size $ L^4 $. The anomaly equation can be written as
\bea
\label{eq:AWI}
\langle \partial_\mu j_\mu^5 (x) \rangle = 2 \anx
     + 2 \sum_{s=1}^{n_{+}} [\phi^s_{+}(x)]^{\dagger} \phi^s_{+} (x)
     - 2 \sum_{t=1}^{n_{-}} [\phi^t_{-}(x)]^{\dagger} \phi^t_{-} (x)
\eea
where $ j_\mu^5 (x) = \bar\psi (x) \gamma_\mu \gamma_5 \psi(x) $
is the axial vector current, $ \langle \partial_\mu j_\mu^5 (x) \rangle $
is the fermionic average of the divergence of axial vector current,
$ \phi^s_{+} $ and $ \phi^t_{-} $ are normalized
eigenfunctions of the zero modes with chirality $ +1 $ and $ -1 $
respectively, and $ \anx $ is the axial anomaly.
Integrating the anomaly equation (\ref{eq:AWI}) over the 4-torus,
the l.h.s. vanishes automatically, and the r.h.s. gives
\bea
\label{eq:anx_index_Q}
\int_{T^4} d^4 x \ \anx = n_{-} - n_{+} \ .
\eea
Using the Atiyah-Singer index theorem (\ref{eq:index_thm}),
Eq. (\ref{eq:anx_index_Q}) becomes
\bea
\label{eq:anx_Q}
\int_{T^4} d^4 x \ \anx = Q = \int d^4 x \ \rho(x) \ ,
\eea
where $ \rho(x) $ is the topological charge density defined in
(\ref{eq:rho}). Then (\ref{eq:anx_Q}) yields
\bea
\label{eq:dg}
\int_{T^4} d^4 x \ ( \anx - \rho(x) ) = 0 \ .
\eea
According to the divergence theorem, (\ref{eq:dg}) implies that
\bea
\label{eq:anxc}
\anx = \rho(x) + \partial_\mu g_\mu (x) \ ,
\eea
where the current $ g_\mu (x) $ is gauge invariant and continuous
on the 4-torus.

For weak gauge backgrounds ( $ | A_\mu (x) | \ll 1/L $ )
with zero topological charge,
$ \anx $ can be evaluated using weak-coupling perturbation theory,
and the result is equal to the topological charge density of the
gauge background,
\bea
\label{eq:anx_p}
\anx = \rho(x) = \frac{1}{32 \pi^2} \epsilon_{\mu\nu\lambda\sigma}
                        \tr ( F_{\mu\nu}(x) F_{\lambda\sigma}(x) ) \ .
\eea
Therefore $ \partial_\mu g_\mu (x) = 0 $ in this case.

Next we consider a nontrivial gauge background with constant
field tensor $ F_{\mu\nu}(x) = F_{\mu\nu}^{0} $.
Since field tensor is constant, $ \anx $ must be constant.
Thus $ \anx $ can be solved directly from Eq. (\ref{eq:anx_Q}),
\bea
\label{eq:anx_rho_c}
\anx = \frac{1}{32 \pi^2} \epsilon_{\mu\nu\lambda\sigma}
       \tr ( F_{\mu\nu}^{0} F_{\lambda\sigma}^{0} ) = \rho_0 \ .
\eea
Again, $ \partial_\mu g_\mu (x) = 0 $ in this case.

It can be shown that $ \partial_\mu g_\mu (x) = 0 $
for any smooth gauge backgrounds.

\subsection{ On a finite lattice }

Next we regularize the theory by discretizing the 4-torus into
a hypercubical lattice with $ N $ sites in each direction
( i.e., $ L = N a $ ). The gauge fields $ \{ A_\mu (x) \} $ are
transcribed to link variables $ \{ U_\mu (x) \} $ on the lattice.
For any lattice Dirac operator $ D $, the anomaly equation on the
finite lattice is \cite{twc99:6},
\bea
\label{eq:AWI_lat}
a^4 \langle \partial_\mu J_\mu^5 (x) \rangle = 2 a^4 \anxL
     + 2 \sum_{s=1}^{n_{+}} [\phi^s_{+}(x)]^{\dagger} \phi^s_{+} (x)
     - 2 \sum_{t=1}^{n_{-}} [\phi^t_{-}(x)]^{\dagger} \phi^t_{-} (x)
\eea
where the axial anomaly is
\bea
\label{eq:anxL}
a^4 \anxL = \lim_{m \to 0 } \frac{1}{4} \ \tr \left[ ( B \hat{D}^{-1} ) (x,x)
                                +( \hat{D}^{-1} B ) (x,x) \right] \ .
\eea
Here $ \hat{D} = D + m $, and $ B $ is an irrelevant operator
denoting the chirality breaking of $ D $,
\bea
\label{eq:DB}
D \gamma_5 + \gamma_5 D = B \ .
\eea
In particular, for GW Dirac operators, $ B = 2 r a D \gamma_5 D $, then
(\ref{eq:anxL}) reduces to \cite{ph98:1,ml98:2,hn97:7}
\bea
\label{eq:anxL_GW}
a^4 \anxL = r \ a \ \tr \left[ \gamma_5 D (x,x) \right] \ .
\eea

Summing the anomaly equation (\ref{eq:AWI_lat}) over
all sites of the finite lattice, then the l.h.s. vanishes and the
r.h.s. gives
\bea
\label{eq:index_anxL}
n_{-} - n_{+} = \sum_{x} a^4 \anxL \ .
\eea
From (\ref{eq:index_anxL}) and (\ref{eq:cD}), we obtain
\bea
\label{eq:anxL_cD}
\sum_{x} a^4 \anxL = c[D] \ Q \ .
\eea

Now we can write
\bea
\label{eq:Q_rhoL}
Q =  \sum_x a^4 \rho_{L} (x) \ ,
\eea
where $ \rho_L (x) $ is the topological charge density of the
gauge background on the finite lattice. Note that we do not
need an explicit expression of $ \rho_L (x) $, which is
supposed to be expressed in terms of link variables.

Then from (\ref{eq:anxL_cD}) and (\ref{eq:Q_rhoL}), we obtain
\bea
\label{eq:anxL_rhoL}
\sum_x ( \anxL - c[D] \rho_L (x) ) = 0 \ .
\eea
The general solution of (\ref{eq:anxL_rhoL}) is
\bea
\label{eq:anxL_rhoL_G}
\anxL = c[D] \rho_L (x) + g(x) \ ,
\eea
where $ g(x) $ satisfies $ \sum_x g(x) = 0 $. We can rewrite
$ g(x) $ in a form analogous to $ \partial_\mu g_\mu (x) $
in (\ref{eq:anxc}),
\bea
\label{eq:div_g}
g(x) = \sum_\mu [ g_\mu (x ) - g_\mu (x - a \hat\mu ) ]
       \equiv \partial_\mu g_\mu (x) \ ,
\eea
where $ g_\mu (x) $ satisfies the periodic boundary conditions :
$ g_\mu ( x + L \hat\nu ) = g_\mu ( x ) $ for $ \mu, \nu = 1, \cdots, 4 $.

Now we write
\bea
\label{eq:rho_av_f}
\rho_L (x) = \bar\rho(x) + f(x)  \ ,
\eea
where $ \bar\rho(x) $ is the topological charge density
inside the unit cell of volume $ a^4 $ centered
at the site $x$, defined in Eq. (\ref{eq:rho_bar}),
and $ f(x) $ denotes the difference between
$ \rho_L(x) $ and $ \bar\rho(x) $.
Summing Eq. (\ref{eq:rho_av_f}) over all sites and using the
relation $ \sum_x a^4 \rho_L(x) = \sum_x a^4 \bar\rho(x) = Q $,
we obtain
\bea
\label{eq:sum_f}
\sum_x f(x) = 0 \ .
\eea
Therefore, we can rewrite $ f(x) $ in terms of a total divergence term
\bea
\label{eq:div_f}
f(x) = \sum_\mu [ f_\mu (x ) - f_\mu (x - a \hat\mu ) ]
       \equiv \partial_\mu f_\mu (x) \ ,
\eea
where $ f_\mu(x) $ satisfies the periodic boundary conditions :
$ f_\mu ( x + L \hat\nu ) = f_\mu ( x ) $
for $ \mu, \nu = 1, \cdots, 4 $.
Substituting (\ref{eq:rho_av_f}) into (\ref{eq:anxL_rhoL_G}), we obtain
a general expression for $ \anxL $,
\bea
\label{eq:anxL_rho_G}
\anxL = c[D] \bar\rho(x) + \partial_\mu G_\mu (x) \ ,
\eea
where
\BAN
\label{eq:G}
\partial_\mu G_\mu (x) = \sum_\mu [ G_\mu (x ) - G_\mu (x - a \hat\mu ) ]
= c[D] f(x) + g(x) \ .
\EAN

If the variation of $ \rho(x) $ is very small across the unit
cell $ a^4 $ centered at $ x $ ( which must be the case in the limit
$ N \to \infty $ ( $ a \to 0 $ ) for a smooth background ),
then $ \bar\rho(x) \simeq \rho(x) $, and $ \anxL $ becomes
\bea
\label{eq:anxLc}
\anxL \simeq c[D] \frac{1}{32 \pi^2}
             \epsilon_{\mu\nu\lambda\sigma}
             \tr ( F_{\mu\nu}(x) F_{\lambda\sigma}(x) )
               + \partial_\mu G_\mu (x) \ .
\eea
where $ \tr $ denotes the trace over the gauge group space.

Equation (\ref{eq:anxL_rho_G}) is a general
expression for the axial anomaly of {\it any} lattice Dirac operator
on a finite lattice. Note that we have not imposed any
conditions on $ D $ except assuming that the exact zero
modes ( if any ) of $ D $ have definite chirality ( this is the case
for GW Dirac operators ) which has been used
in deriving the anomaly equation (\ref{eq:AWI_lat}).

From (\ref{eq:anxLc}), we immediately see that $ \anxL $ can
recover the continuum axial anomaly if and only if $ c[D]=1 $
and $ \partial_\mu G_\mu (x) = 0 $ for all $ x $.
However, satisfying the five conditions {\bf (i)-(v)} does not
guarantee that $ c[D] = 1 $ and $ \partial_\mu G_\mu (x) = 0 $
in nontrivial gauge backgrounds.

Next we try to understand the total divergence term
$ \partial_\mu G_\mu (x) $.
Let us consider a nontrivial gauge configuration with
constant field tensor. Then the topological charge density is
constant and the topological charge is,
\bea
Q &=& \sum_x a^4 \rho_L (x) = N^4 a^4 \rho_0 \ , \nn
\label{eq:rho_c}
\rho_0 &=& \frac{1}{32 \pi^2} \epsilon_{\mu\nu\lambda\sigma}
                     \tr ( F_{\mu\nu}^{0} F_{\lambda\sigma}^{0} ) \ .
\eea
An example is given in Eqs. (\ref{eq:U1a})-(\ref{eq:U4a}).

Since $ \rho_L(x) $ is constant, one may expect that
$ \anxL $ is constant too. However, this is true
only if $ D $ is local. In general, if $ D $ is
local in the free fermion limit, then it would be local in a gauge
background which satisfies the locality bound (\ref{eq:U_bound_2}).
If we assume that is the case, then $ \anxL $ is constant and it
can be solved directly from Eq. (\ref{eq:anxL_cD}),
\bea
\label{eq:anxL_rho_c}
\anxL = c[D] \rho_L (x) = c[D] \rho_0 \ .
\eea
Comparing (\ref{eq:anxL_rho_c}) to (\ref{eq:anxLc}),
we obtain $ \partial_\mu G_\mu (x) = 0 $ provided that $ c[D] \ne 0 $.
If $ c[D] = 0 $, then Eq. (\ref{eq:anxL_cD}) becomes
\bea
\label{eq:sum_anxL_0}
\sum_{x} \anxL = 0 \ ,
\eea
which implies that $ \anxL = 0 $ if $ \anxL $ is constant.
However, in a gauge background with nonzero constant field
tensor, $ \anxL $ cannot be zero at all sites,
since $ D $ must at least have some responses to the gauge
background except for $ D \equiv 0 $.
Therefore, in the case $ c[D] = 0 $, $ \anxL $ cannot be identically
zero. Thus (\ref{eq:sum_anxL_0}) implies that $ \anxL $ can be
written as a total divergence,
\bea
\anxL = \partial_\mu g_\mu (x) \ .
\eea
Therefore, in a nontrivial gauge background with constant field tensor,
the axial anomaly on a finite lattice is
\bea
\label{eq:anxL_const}
\anxL = \left\{ \begin{array}{cc}
                 c[D] \rho_0 \ ,            & \hspace{4mm}  c[D] \ne 0   \\
                 \partial_\mu g_\mu (x) \ , & \hspace{4mm}  c[D] = 0
               \end{array}  \right.
\eea
Thus a topologically trivial $ D $ cannot have the correct
axial anomaly in nontrivial gauge sectors.

Now suppose $ D $ is topologically proper.
If we introduce very small local fluctuations of $ \rho(x) $
on top of the constant $ \rho_0 $ such that $ \sum_x \delta \rho(x) = 0 $,
then evidently the axial anomaly of $ D $ becomes
\bea
\label{eq:anxL_rho}
\anxL = \bar\rho(x) \ .
\eea
Further, one can deduce that Eq. (\ref{eq:anxL_rho}) holds for any gauge
configurations satisfying the locality bound (\ref{eq:U_bound_2}).

In the trivial sector ( $ Q = 0 $ ), for weak gauge backgrounds
( $ | a A_\mu (x) | << 1/N $ ) at finite lattice spacing
or smooth backgrounds in the classical continuum limit,
the axial anomaly $ \anxL $ can be evaluated using perturbation theory
( For examples, see ref. \cite{kiku98:6}-\cite{tr99:3} ).
If $ D $ satisfies the five conditions {\bf (i)-(v)}, then
$ \anxL $ is equal to the topological charge density of the
gauge background,
\bea
\label{eq:anxL_p}
\anxL = \frac{1}{32 \pi^2} \epsilon_{\mu\nu\lambda\sigma}
                  \tr ( F_{\mu\nu}(x) F_{\lambda\sigma}(x) ) \ .
\eea
Thus even if $ D $ is topologically trivial in nontrivial gauge
backgrounds, it can have correct axial anomaly in the trivial
gauge sector.

\subsection{Examples}

In the following, we illustrate the results of (\ref{eq:anxL_const})
and (\ref{eq:anxL_p}) by explicit examples.

First, for the purpose of comparison, we also consider the
Wilson-Dirac lattice fermion operator \cite{wilson75} which does not
satisfy the Ginsparg-Wilson relation {\bf (v)} but the other four
conditions {\bf (i)-(iv)}.
Explicitly, the massless Wilson-Dirac operator can be written as
\bea
\label{eq:Dw}
D_W = \gamma_\mu t_\mu + W \ ,
\eea
where $ t_\mu $ and $ W $ are defined in Eqs. (\ref{eq:tmu}) and
(\ref{eq:wilson}) respectively. According to formula (\ref{eq:anxL}),
the axial anomaly of $ D_W $ can be written
\bea
\label{eq:anxw}
\anxw &=&  \frac{1}{4} \sum_{\mu} \Bigl(
    4 \ \tr [ D_w^{-1} (x,x) \gm5 ]
    - \ \tr [ D_w^{-1}(x,x+a\hat\mu) \gm5 U_{\mu}(x) ]  \nn
&-& \tr [ D_w^{-1}(x+a\hat\mu,x) \gm5 U_{\mu}^{\dagger}(x) ]
   - \ \tr [ D_w^{-1}(x-a\hat\mu,x) \gm5 U_{\mu}(x-a\hat\mu) ] \nn
&-& \tr [ D_w^{-1}(x,x-a\hat\mu) \gm5 U_{\mu}^{\dagger}(x-a\hat\mu)] \ \Bigr).
\eea
where $ \tr $ denotes the traces over the Dirac space and the gauge
group space. It is well known that $ D_W $ does not possess any exact
zero modes in topologically nontrivial gauge fields, thus $ c[D_W] = 0 $.
Therefore its axial anomaly must disagree with the topological charge
density in nontrivial gauge backgrounds, according to (\ref{eq:anxLc}).
However, one may wonder whether $ D_W $ can produce the correct axial
anomaly in trivial gauge backgrounds satisfying the locality bound.
For this purpose, it suffices to consider the trivial $ U(1) $ gauge
backgrounds with local sinusoidal fluctuations on a 2-dimensional
flat torus of size $ L \times L $ :
\bea
\label{eq:A1_2d}
a A_1(x) &=& \frac{ 2 \pi h_1 a }{L} +
             A_1^{(0)} \sin \left( \frac{ 2 \pi n_2 }{L} x_2 \right) \\
\label{eq:A2_2d}
a A_2(x) &=& \frac{ 2 \pi h_2 a }{L} +
             A_2^{(0)} \sin \left( \frac{ 2 \pi n_1 }{L} x_1 \right)
\eea
The corresponding link variables on the 2d lattice
( $ x_\mu = 0, a, ..., ( N - 1 ) a $ ) are
\BAN
U_1(x) &=& \exp \left[ \ \text{i} a A_1(x) \ \right] \ , \\
U_2(x) &=& \exp \left[ \ \text{i} a A_2(x) \ \right] \ .
\EAN
The topological charge density on the 2d torus is
\bea
\label{eq:rho_2d}
\rho (x) &=& \frac{1}{2\pi} F_{12} (x)  \nn
         &=&
     A_2^{(0)} \frac{n_1}{a L} \cos \left( \frac{ 2 \pi n_1 }{L} x_1 \right)
  -  A_1^{(0)} \frac{n_2}{a L} \cos \left( \frac{ 2 \pi n_2 }{L} x_2 \right)
\eea
The topological charge density inside the square $ a^2 $
centered at $ x $ is
\bea
\label{eq:rho_bar_2d}
\hspace{-4mm} & & \hspace{-4mm}
\bar\rho(x) = \frac{1}{a^2} \int_{x_1 - a/2}^{x_1 + a/2 } d x_1
              \int_{x_2 - a/2}^{x_2 + a/2 } d x_2 \ \rho(x) \\
\hspace{-4mm} &=& \hspace{-4mm}
\frac{1}{a^2 \pi} \left[ A_2^{(0)} \sin \left( \frac{ \pi n_1 a }{L} \right)
                  \cos \left( \frac{ 2 \pi n_1 }{L} x_1 \right)
                  -A_1^{(0)} \sin \left( \frac{\pi n_2 a}{L} \right)
                  \cos \left( \frac{ 2 \pi n_2 }{L} x_2 \right) \right]
\eea

In Fig. \ref{fig:anxw}, we plot $ \anxw $ for each site on a
$ 12 \times 12 $ lattice with lattice spacing $ a = 1 $,
comparing with the topological charge density $ \bar\rho(x) $, in a
trivial gauge background [ Eqs. (\ref{eq:A1_2d}) and (\ref{eq:A2_2d}) ]
with parameters $ A_1^{(0)} = 0.3 $,
$ A_2^{(0)} = 0.4 $, $ n_1 = n_2 = 1 $,
$ h_1 = 0.1 $ and $ h_2 = 0.2 $.
Note that in this case, the difference between $ \rho(x) $ and
$ \bar\rho(x) $ is very small, always less than one percent.
The position of a site with coordinates $ ( x_1, x_2 ) $ is represented
by an integer $ x = 12 ( x_2 - 1 ) + x_1 $, as the x-coordinate in Fig. 1.
The axial anomaly $ \anxw $ is denoted by squares. The topological charge
density $ \bar\rho(x) $ (\ref{eq:rho_bar_2d}) of the gauge background is
denoted by circles.
The line segments between circles are inserted only for the visual purpose.
Evidently $ \anxw $ disagrees with the topological charge
density $ \bar\rho(x) $.
The discrepancies are due to the presence of the
fermion doublers which decouple completely {\it only} in the limit
$ a \to 0 $.

The deviation of the axial anomaly
of a lattice Dirac operator in a gauge background
can be measured in terms of
\bea
\label{eq:delta}
\delta = \frac{1}{N_s} \sum_x \frac{| \anxL - \bar\rho(x)|}{|\bar\rho(x)|}
\eea
where $ N_s $ is the total number of sites of the lattice,
and $ \bar\rho(x) $ is the topological charge density inside
the unit cell of volume $ a^d $ centered at $ x $.

For the axial anomaly of the Wilson-Dirac operator as
shown in Fig. \ref{fig:anxw}, the deviation is $ \delta_W = 0.45 $.
One expects that the deviation $ \delta_W $ goes to zero only in the
limit $ N \to \infty $ ( or $ a \to 0 $ ).

\psfigure 5.0in -0.2in {fig:anxw} {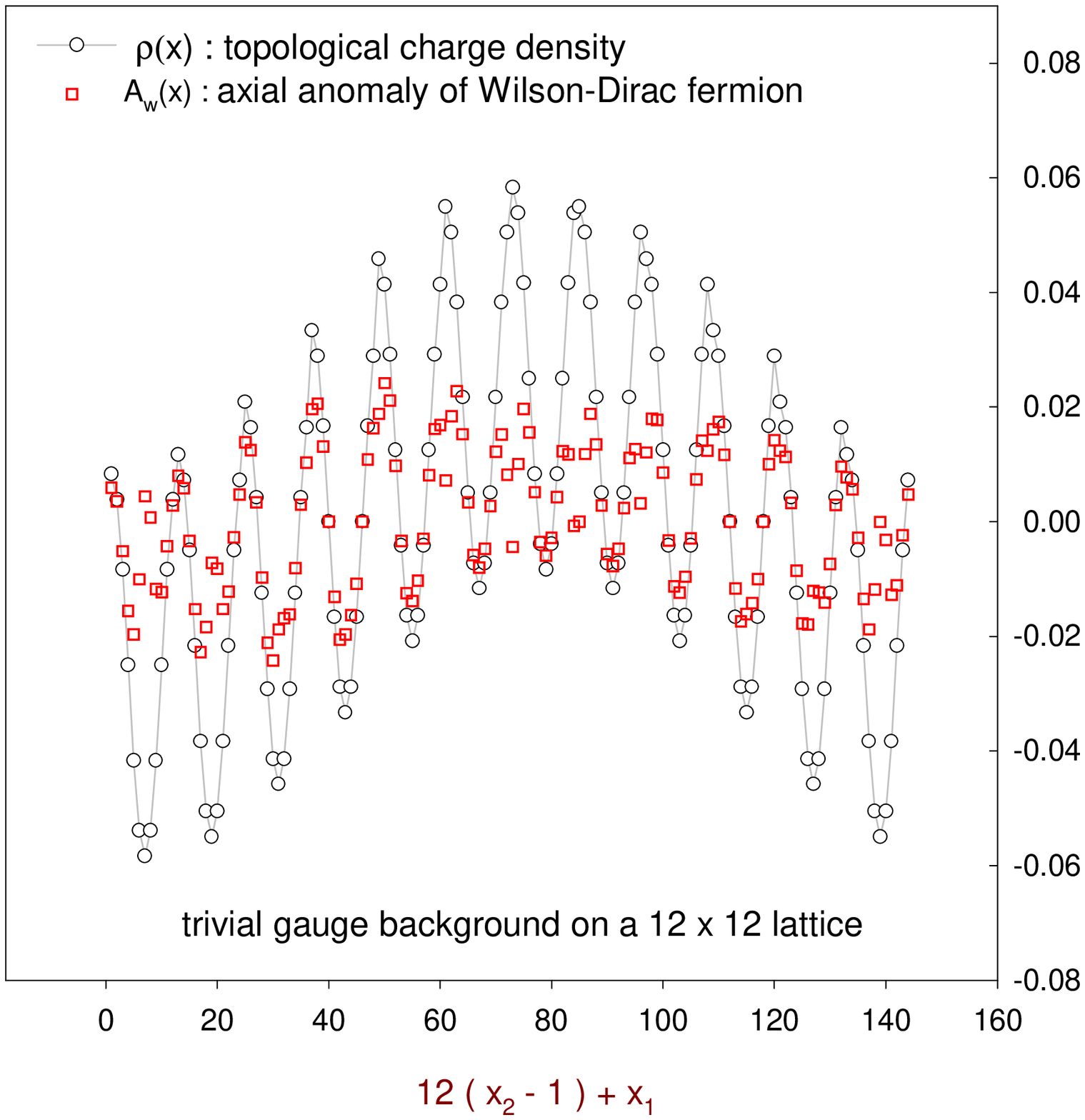} {
The axial anomaly $ \anxw $ [ Eq. (\ref{eq:anxw}) ] of the massless
Wilson-Dirac operator $ D_w $ [ Eq. (\ref{eq:Dw}) ] in a trivial
gauge background on a $ 12 \times 12 $ lattice.
The background $ U(1) $ gauge field
[ Eqs. (\ref{eq:A1_2d})-(\ref{eq:A2_2d}) ]
is specified by the parameters $ h_1 = 0.1 $, $ h_2 = 0.2 $,
$ A_1^{(0)} = 0.3 $, $ A_2^{(0)} = 0.4 $ and $ n_1 = n_2 = 1 $.
The axial anomaly of the Wilson-Dirac operator, $ \anxw $, is denoted by
squares. The topological charge density $ \bar\rho(x) $
[ Eq. (\ref{eq:rho_bar_2d}) ]
of the gauge background is denoted by circles.
The line segments between circles are inserted only for the visual purpose.
Evidently $ \anxw $ disagrees with $ \bar\rho(x) $.
}


Next we consider the lattice Dirac operators
[ (\ref{eq:Dtwc}) and (\ref{eq:Dw_GW}) ] in Section 3,
which satisfy all five conditions {\bf (i)-(v)}, but both are
topologically trivial. According to (\ref{eq:anxL}), the axial
anomaly of (\ref{eq:Dtwc}) is
\bea
\label{eq:anxt}
\anxt = r^2 \ \tr \left[ ( b b^{\dagger} + r^2 )^{-1} -
                         ( b^{\dagger} b + r^2 )^{-1} \right] (x,x) \ ,
\eea
and that of (\ref{eq:Dw_GW}) is
\bea
\label{eq:anxwc}
\anxwc =  r^2 \ \tr \left[ C^{\dagger} C ( \Id + r^2 C^{\dagger} C )^{-1}
            - C C^{\dagger} ( \Id + r^2 C C^{\dagger} )^{-1} \right] (x,x) \ ,
\eea
where $ \tr $ denotes the traces over the Dirac space
and the gauge group space.
The axial anomalies $ \anxt $ and $ \anxwc $ are plotted in
Fig. \ref{fig:anxt} and Fig. \ref{fig:anxwc} respectively,
for the same gauge background as in Fig. \ref{fig:anxw}.
The deviation for the axial anomaly $ \anxt $ shown in Fig. \ref{fig:anxt}
is $ \delta = 0.121 $, and for $ \anxwc $ shown in Fig. \ref{fig:anxwc} is
$ \delta = 0.068 $. ( For comparison, the deviation for the
axial anomaly of the Neuberger-Dirac operator with $ m_0 = 1 $ is
$ \delta = 0.049 $ for the same gauge background in Fig. \ref{fig:anxw} ).
Evidently both $ \anxt $ and $ \anxwc $ agree with
$ \bar\rho(x) $ ( $ \rho(x) $ ) for the trivial gauge background.
They provide explicit demonstrations of (\ref{eq:anxL_p}) which holds
for any $ D $ satisfying the five conditions {\bf (i)-(v)}, and for any
trivial gauge backgrounds satisfying the locality bound.
It is instructive to compare the axial anomalies of the GW Dirac operators
in Fig. \ref{fig:anxt} and Fig. \ref{fig:anxwc}
to that of the Wilson-Dirac operator in Fig. \ref{fig:anxw}.
The former ones can reproduce the continuum axial anomaly
even on a finite lattice while the latter cannot.

In Fig. \ref{fig:anx12}, we plot the axial
anomaly in a nontrivial gauge background
( $ Q = 2 $ ) with constant field strength on a
$ 12 \times 12 $ lattice, for
the topologically proper Neuberger-Dirac operator, as well as
the topologically improper GW Dirac operators (\ref{eq:Dtwc}) and
(\ref{eq:Dw_GW}), respectively. It is clear that the axial anomaly of
the Neuberger-Dirac operator agrees with the constant topological
charge density $ \rho_0 = 1/72 $ at each site, while those of
the GW Dirac operators (\ref{eq:Dtwc}) and (\ref{eq:Dw_GW})
are in complete disagreement with $ \rho_0 $. This provides
an explicit demonstration of Eq. (\ref{eq:anxL_const}) for
two different cases $ c[D]=1 $ and $ c[D]=0 $. For the case $ c[D]=0 $,
it shows explicitly that the axial anomaly is in the form
$ \partial_\mu g_\mu (x) $, which oscillates around zero
with its sum over all sites equal to zero.

\psfigure 5.0in -0.2in {fig:anxt} {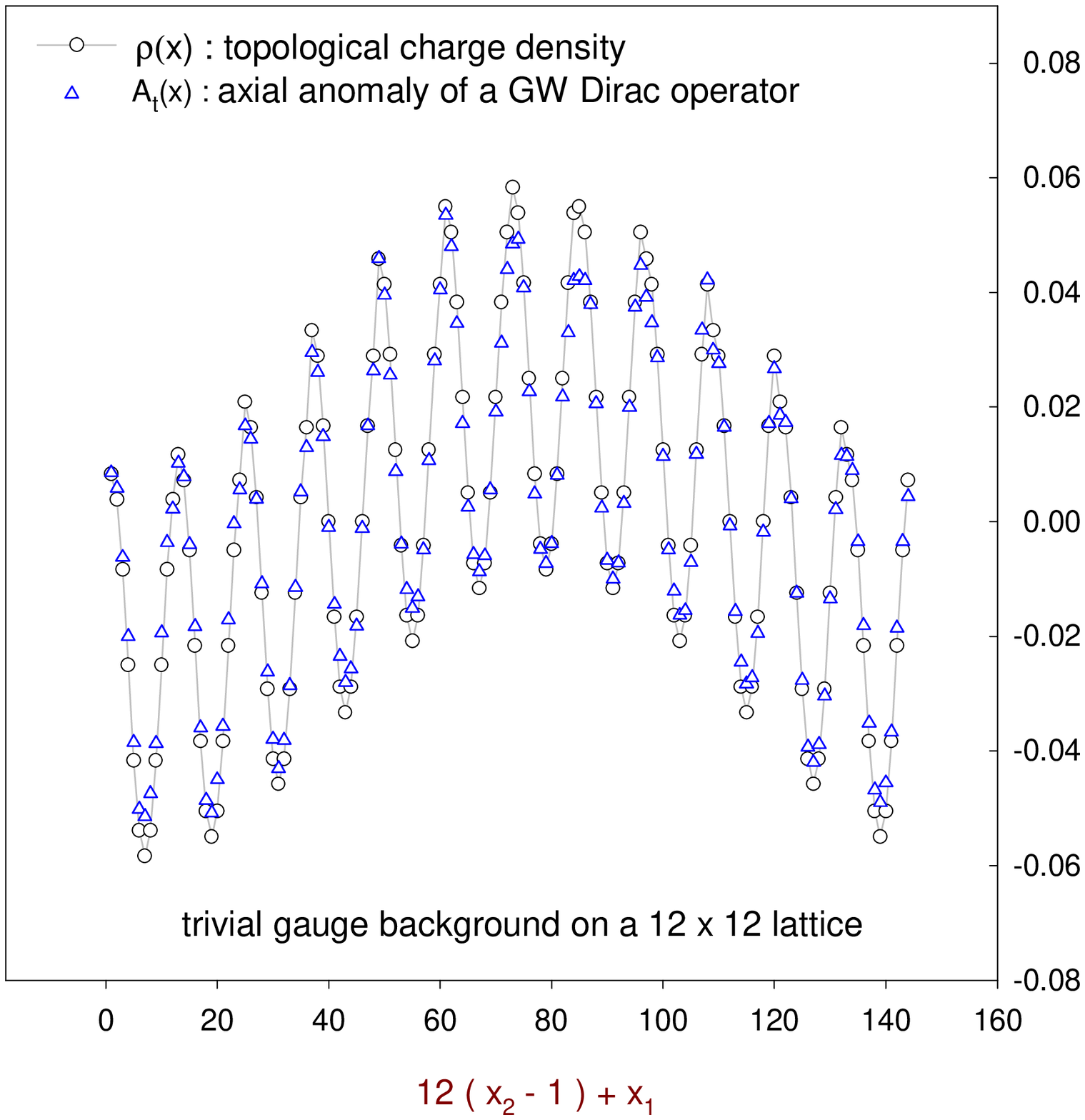} {
The axial anomaly $ \anxt $ [ Eq. (\ref{eq:anxt}) ] of the massless
GW Dirac operator $ D $ [ Eq. (\ref{eq:Dtwc}) ] in a trivial
gauge background on a $ 12 \times 12 $ lattice.
The value of $ r $ in $ D $ has been set to $ 0.4 $, and there is
no significant changes to $ \anxt $ for any $ r $ in the range
$ \sim 0.2 $ to $ \sim 0.8 $.
The background $ U(1) $ gauge field is the same as that in Fig. 1.
The axial anomaly $ \anxt $ is denoted by triangles.
The topological charge density $ \bar\rho(x) $ [ Eq. (\ref{eq:rho_bar_2d}) ]
of the gauge background is denoted by circles.
The line segments between circles are inserted only for the visual purpose.
}

\psfigure 5.0in -0.2in {fig:anxwc} {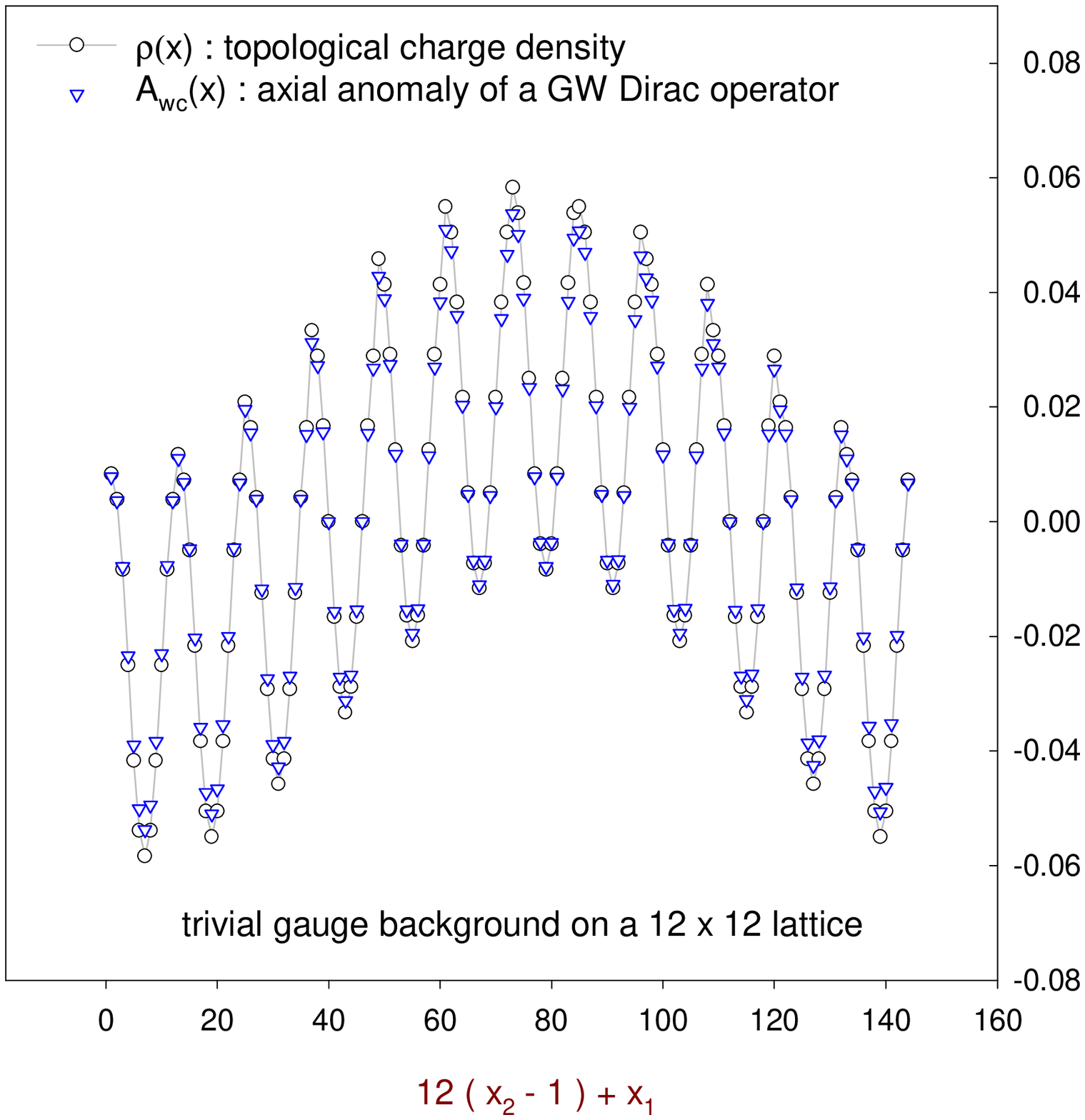} {
The axial anomaly $ \anxwc $ [ Eq. (\ref{eq:anxwc}) ] of the massless
GW Dirac operator $ D $ [ Eq. (\ref{eq:Dw_GW}) ] in a trivial
gauge background on a $ 12 \times 12 $ lattice.
The value of $ r $ in $ D $ has been set to $ 0.5 $, and there is
no significant changes to $ \anxwc $ for any $ r $ in the range
$ \sim 0.2 $ to $ \sim 0.8 $.
The background $ U(1) $ gauge field is the same as that in Fig. 1.
The axial anomaly $ \anxwc $ is denoted by triangles.
The topological charge density $ \bar\rho(x) $ [ Eq. (\ref{eq:rho_bar_2d}) ]
of the gauge background is denoted by circles.
The line segments between circles are inserted only for the visual purpose.
}

\psfigure 5.0in -0.2in {fig:anx12} {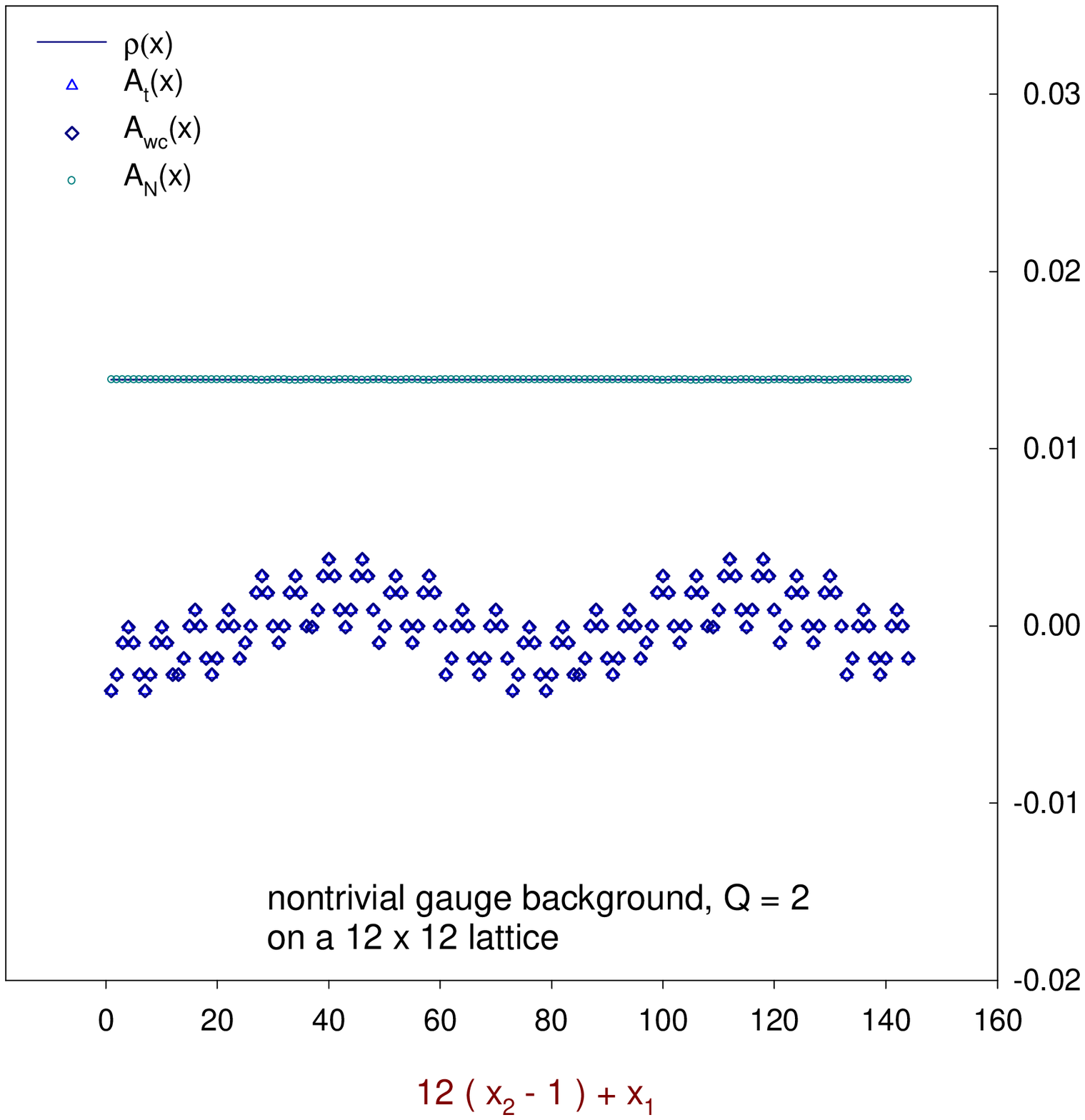} {
The axial anomalies of GW Dirac operators in a nontrivial
gauge background with constant field tensor $ F_{12} = \pi/36 $
( topological charge $ Q = 2 $ ) on a $ 12 \times 12 $ lattice.
The topological charge density ( $ \rho(x) = 1/72 $ )
of the gauge background is denoted by the horizontal line.
The axial anomaly $ {\cal A}_N(x) $ of the Neuberger-Dirac operator
agrees with $ 1/72 $ excellently, lying on the horizontal line.
The axial anomaly $ \anxwc $ of (\ref{eq:Dw_GW})
is denoted by diamonds, while $ \anxt $ of (\ref{eq:Dtwc}) is denoted
by triangles. Evidently $ \anxwc $ and $ \anxt $ both
disagree with the topological charge density $ \rho(x) = 1/72 $.
}

In passing, we note that the topological bound of the
Neuberger-Dirac operator with $ m_0 = 1 $ in two dimensions is
$$
a^2 | \bar\rho(x) | < \epsilon_1 \simeq 0.28 \hspace{4mm} \forall \ x \ ,
$$
while the locality bound \cite{hn99:11a} is
$$
|| \Id - U_{\mu\nu}(x) || <
\epsilon = \frac{1}{2 + \sqrt{2} } \simeq 0.29 \hspace{2mm}
\mbox{ for all plaquettes } \ ,
$$
which is transcribed into
$$
a^2 | \bar\rho(x) | < \epsilon_2 \simeq 0.046 \hspace{4mm} \forall \ x \ .
$$
Again, $ \epsilon_2 $ is much less than $ \epsilon_1 $.

\section{Summary and Discussions}

Given any lattice Dirac operator $ D_0 $
satisfying the four conditions {\bf (i)-(iv)}, one can construct a
chirally symmetric lattice Dirac operator $ D_c $ satisfying the three
conditions {\bf (ii)-(iv)}, i.e.,
\bea
\label{eq:Dc_D0}
D_c = 2 \gm5 D_0 ( \gm5 D_0 - D_0 \gm5 )^{-1} D_0 \ ,
\eea
which must be nonlocal as a consequence of the Nielson-Ninomiya
theorem \cite{no-go}.
An example of such $ D_c $ is given in Eq. (\ref{eq:Dwc}).
Then the GW Dirac operator $ D = D_c ( \Id + r a D_c )^{-1} $
satisfies all five conditions {\bf (i)-(v)}, where
the locality of $ D $ is ensured by choosing $ r $ in the proper range.
It is obvious that one can replace (\ref{eq:Dc_D0}) by any
$ D_c $ which satisfies the three conditions {\bf (ii)-(iv)}.
An example is given in Eqs. (\ref{eq:Dc1})-(\ref{eq:D_L}).

If $ D_c $ is well-defined ( without any singularities )
in nontrivial gauge backgrounds, then the index of
$ D = D_c ( \Id + r a D_c )^{-1} $ is zero \cite{twc98:6a}.
In other words, even if $ D $ satisfies all
five conditions {\bf (i)-(v)}, $ D $ does {\it not} necessarily have
exact zero modes in nontrivial gauge fields.
Two explicit examples have been presented in Section 3.
At present, we do not know how to formulate additional constraints
( in useful forms ) such that $ D $ can produce exact zero modes
satisfying the Atiyah-Singer index theorem.
Even though such conditions will be found in the future,
it is still convenient for us to introduce the topological
characteristics $ c[D] $ associated with each lattice Dirac operator,
as defined in (\ref{eq:cD_def}).
For a given $ D $, $ c[D] $ is an integer constant provided that
the topological charge inside any unit cell is small enough,
i.e., $ a^4 | \bar\rho(x) | < \epsilon_1 $.
( For Neuberger-Dirac operator with $ m_0 = 1 $, $ \epsilon_1 \simeq 0.02 $ ).
However, in general, one does {\it not} need to go to the limit $ a = 0 $
( or $ N = \infty $ ) in order to reveal the topological
characteristics of any $ D $.
Our view is that if a lattice Dirac operator
( satisfying {\bf (i)-(v)} ) does not have the correct index
on {\it any} finite lattices, then it will {\it never} have the
correct index in the continuum limit.
An argument has been presented in Section 2.

Since the index of $ D = D_c ( \Id + r a D_c )^{-1} $ is invariant
for any $ r $, the topological characteristics $ c[D] $ is also
invariant from $ r = 0 $, $ D = D_c $ ( a nonlocal operator ) up to
$ r a \gg 1 $,
$ D \simeq ( r a )^{-1} \Id + \mbox{ higher order corrections } $
( a highly local operator ). Thus the locality condition {\bf (i)}
does not form a constraint to $ c[D] $, even though it is crucial
to recover the continuum axial anomaly on the lattice when $ c[D] = 1 $.
Further, one can always transform any lattice Dirac operator into
a GW Dirac operator with the same index \cite{twc99:12a}, so
the GW relation {\bf (v)} also does not form a constraint
to $ c[D] $. Therefore, we conclude that among these five conditions
{\bf (i)-(v)}, only three of them
[ i.e., {\bf (ii), (iii)} and {\bf (iv)} ] provide constraints to $ c[D] $.
However, satisfying all of them does not guarantee that $ c[D] = 1 $.
Additional constraints are needed. They should be formulated
explicitly with exact solutions rather than a set of
equations with only implicit or approximate solutions.
From this viewpoint, the Neuberger-Dirac operator is {\it much more}
than just a solution of the GW relation, since the former has the
proper topological characteristics, while the latter
does not necessarily have. This seems to indicate that
the Overlap formalism \cite{rn95,sr95}
( and the Domain-Wall fermion \cite{kaplan92} ) indeed plays
the fundamental role in solving the problem of chiral fermions
on the lattice.

It is instructive to compare L\"uscher's definition of
topological charge \cite{ml82} to that defined by the index of
the Neuberger-Dirac operator. The former definition
relies on the fact that if the gauge field
satisfies the bound
\bea
\label{eq:LB}
| \tr ( \Id - U_{\mu\nu}(x) ) | < \epsilon  \hspace{2mm}
\mbox{ for all plaquettes, }
\eea
then any link configuration ( with nonzero integer topological charge )
cannot deform into the trivial configuration
( with all link variables equal to the identity ), and vice versa.
Thus the gauge link configurations can be decomposed into disconnected
topological sectors as in the continuum, and the topological charge
is a well-defined integer even on a lattice.
For gauge fields satisfying the bound (\ref{eq:LB}), L\"uscher derived
an expression for the topological charge density on the lattice
$ \rho_L (x) $ ( i.e., $ a^{-4} q(n) $ of Eq. (32) in ref. \cite{ml82} ),
which satisfies $ Q = \sum_x a^4 \rho_L(x) $, and it tends to
$ \rho(x) $ {\it smoothly} in the continuum limit.
That is, even on a finite lattice, $ \rho_L(x) $ is already
in good agreement with $ \rho(x) $.
On the other hand, the axial anomaly of a topologically proper $ D $
( assuming it is local in the free fermion limit ) is
$ \anxL = \bar\rho(x) \simeq \rho(x) $
provided that $ D $ is local, i.e.,
the gauge background satisfies the locality bound (\ref{eq:U_bound_2}).
Since $ \anxL \simeq \rho_L (x) $ for a topologically proper $ D $
[ see Eq. (\ref{eq:anxL_rhoL_G}) with $ c[D]=1 $, and $ g(x) \simeq 0 $ ],
it follows that the locality bound (\ref{eq:U_bound_2}) of the
Neuberger-Dirac operator is compatible with
L\"uscher's bound (\ref{eq:LB}).
However, if one is only interested in the topological charge,
then the index of the Neuberger-Dirac operator is already equal to
the topological charge for gauge configurations satisfying
the topological bound (\ref{eq:topo_bound}) or (\ref{eq:U_bound_1})
which is less restrictive than the locality bound (\ref{eq:U_bound_2}).
This seems to suggest that if one uses the index of the Neuberger-Dirac
operator \cite{hn97:7} to define the topological charge of a gauge
background, then the lattice size $ N $ can be smaller
( or the lattice spacing $ a $ can be larger )
than that using L\"uscher's definition \cite{ml82}.

A general expression (\ref{eq:anxL_rho_G}) for
the axial anomaly of any lattice Dirac operator has been derived,
in which the role of topological characteristics is displayed explicitly.
For a topologically proper $ D $, its axial anomaly is equal to the
topological charge density $ \rho(x) $ for smooth gauge background
satisfying the locality bound.
Thus, lattice QCD with the Neuberger-Dirac quarks
has the correct axial anomaly associated with the global
chiral symmetry. For chiral gauge theories such as the
standard model, the gauge anomaly must exactly cancel
at finite lattice spacing before one is sure
the existence of a nonperturbative regularization of
the theory. For an anomaly-free fermion multiplet,
it has been argued \cite{ml99:4} that the nonabelian gauge
anomaly cancellation ( in the trivial sector ) holds to
all orders of an expansion in powers of the lattice spacing, while the
abelian gauge anomaly cancellation \cite{ml98:8b, ml98:11} is {\it exact}
at finite lattice spacing. In continuum, the non-abelian gauge anomaly
in $4$-dimensional space may be obtained from the axial anomaly
in $6$-dimensional space \cite{rs84,bz84,AG84}. Thus its form
and normalization follow from the Atiyah-Singer index theorem
for a certain $6$-dimensional Dirac operator.
Presumably, a similar analysis can be performed on a finite
lattice with the axial anomaly in the form of
(\ref{eq:anxL_rho}) for a topologically proper $ D $ in a gauge
background satisfying the locality bound (\ref{eq:U_bound_2}).
It seems that the gauge anomaly can exactly cancel at finite lattice
spacing. We intend to return to this question in a later publication.

To conclude, we emphasize that $ c[D]=1 $ is a stringent requirement
for any construction of lattice Dirac operator $ D $ in $2n$-dimensional
space. It seems that this requirement could not be
satisfied by any approximate solutions to the GW relation plus
other physical constraints in the $2n$-dimensional space.
Unless such $2n$-dimensional $ D $ can possess
{\it exact zero modes and the correct index}
in nontrivial gauge backgrounds on {\it finite} lattices, otherwise
its axial anomaly will never recover the correct result in the
continuum limit.

\bigskip
\bigskip

\appendix

\section{ Some basic properties of GW Dirac operators }

In this appendix, we review some basic properties of the GW Dirac
operator, which are pertinent to our discussions in this paper.
Most of these properties can be found in
refs. \cite{twc98:4, twc98:6a}.
Here we assume that $ D $ satisfies the five conditions {\bf (i)-(v)}
listed in Section 1.

The general solution to the GW relation {\bf (v)} can be
written as \cite{twc98:6a}
\bea
\label{eq:gen_sol}
D = D_c ( \Id + a r D_c )^{-1} = ( \Id + a r D_c )^{-1} D_c
\eea
where $ D_c $ is any chirally symmetric ( $ D_c \gm5 + \gm5 D_c = 0 $ )
Dirac operator which must violate at least one of the three conditions
{\bf (i)-(iii)} above, according to the Nielson-Ninomiya no-go theorem.
Now we require $ D_c $ to satisfy {\bf (ii)} and {\bf (iii)},
but violate {\bf (i)} (  i.e, $ D_c $ is nonlocal ),
since (\ref{eq:gen_sol}) can transform the nonlocal $ D_c $ into a
local $ D $ on a finite lattice for $ r $ in
the proper range, while the properties
{\bf (ii)-(iv)} are preserved. Then $ D $ satisfies all five
conditions {\bf (i)-(v)}.

Moreover, the zero modes and the index of $ D_c $ are invariant under
the transformation (\ref{eq:gen_sol}) \cite{twc98:6a}.
That is, a zero mode of $ D_c $ is also a zero mode of $ D $
and vice versa, i.e.,
\BAN
  D \phi_{\pm} = 0 \Leftrightarrow  D_c \phi_{\pm} = 0,
\EAN
where $ \gamma_5 \phi_{\pm} = \pm \phi_\pm $. Then the number of zero
modes for each chirality must be the same for $ D $ and $ D_c $,
thus the index of the {\it local} $ D $ is equal to the index
of the {\it nonlocal} $ D_c $,
\beq
\label{eq:npm}
 n_{+} ( D_c ) = n_{+} ( D ), \hspace{4mm} n_{-} ( D_c ) = n_{-} ( D ),
\eeq
\beq
\label{eq:index}
\mbox{index}(D_c) = n_{-}(D_c) - n_{+}(D_c) =
 n_{-}(D) - n_{+}(D) = \mbox{index}(D) \ .
\eeq
Therefore, a {\it nonlocal} Dirac operator can have well-defined
index, at least for those obtained by the topologically invariant
transformation
\bea
\label{eq:Dc}
D_c = D ( \Id - a r D )^{-1} = ( \Id - a r D )^{-1} D \ ,
\eea
which is the inverse transform of (\ref{eq:gen_sol}).
From the definition of the topological characteristics $ c[D] $,
(\ref{eq:cD}), we have
\BAN
c[D] = c[D_c] \ .
\EAN

The $\gm5$-hermiticity of $ D $ {\bf (iv)} is equivalent to the
$\gm5$-hermiticity of $ D_c $,
\bea
\label{eq:hermit}
 D_c^{\dagger} = \gm5 D_c \gm5 \ .
\eea
Then the chiral symmetry of $ D_c $ together with its
$\gamma_5$-hermiticity implies that $ D_c $ is
antihermitian ( $ D_c^{\dagger} = -D_c $ ).
This last property is in agreement with the massless Dirac
fermion operator in continuum. Then there exists one to one
correspondence between $ D_c $ and a unitary operator $ V $
such that
\bea
D_c = M (\Id + V )(\Id - V )^{-1}, \hspace{4mm}
V = (D_c - M)( D_c + M)^{-1}.
\label{eq:VDc}
\eea
where $ M $ is a mass scale and $ V $ also satisfies the
$\gm5$-hermiticity $ V^{\dagger} = \gamma_5 V \gamma_5 $.
Then the general solution (\ref{eq:gen_sol}) can be written as
\cite{twc98:6a}
\bea
\label{eq:gwsola}
D = M ( \Id + V )[ ( \Id - V ) + r M a ( \Id + V ) ]^{-1}
\eea

Due to the $\gamma_5$-hermiticity, the eigenvalues of $ D $ are
either real or come in complex conjugate pairs.
Furthermore, from (\ref{eq:gen_sol}), since $ r $ is a positive
real number, the lower bound of real eigenvalues of $ D $
is zero, thus $ \det(D) $ is real and nonnegative, and is amenable
to Hybrid Monte Carlo simulation with {\it any} number of flavors
of dynamical fermions. For the GW relation {\bf (v)} with any
$ r $ ( not restricted to $ 1/2 $ ), the analysis in ref. \cite{twc98:4}
[ Eqs. (24)-(41) ] goes through with trivial modification.
The main results are :
\begin{description}
\item[($\alpha$)] The eigenvalues of $ D $ fall on a circle with center
at $ 1/2r $, and radius $ 1/2r $, and have the reflection symmetry with
respect to the real axis.
\item[($\beta$)] The real eigenmodes ( if any ) at $ 0 $ and $ 1/r $ have
                 definite chirality $ +1 $ or $ -1 $.
\item[($\gamma$)] The chirality of any complex eigenmodes is zero.
\item[($\delta$)] Total chirality of all eigenmodes must vanish. \newline
  $ \Tr ( \gamma_5 ) = \sum_{s} \phi_s^{\dagger} \gamma_5 \phi_s
                     = n_{+} - n_{-} + N_{+} - N_{-} = 0 $
\end{description}
where $ n_{+} ( n_{-} ) $ denotes the number of zero modes of
positive ( negative ) chirality, and $ N_{+} ( N_{-} ) $ the number of
$ 1/r $ modes of positive ( negative ) chirality.
From ($\delta$), we immediately see that {\it any zero mode must be
accompanied by a real } $ 1/r $ {\it mode with opposite chirality},
and the index of $ D $ is
\bea
\label{eq:index_D}
\mbox{index}(D) \equiv n_{-} - n_{+} = - ( N_{-} - N_{+} ) \ .
\eea

Now the central problem is to construct the chirally symmetric
$ D_c $ which is nonlocal, and satisfies {\bf (iii)}, {\bf (iv)},
and (\ref{eq:hermit}). Furthermore we also require that
$ D_c $ is topologically proper ( i.e., satisfying
the Atiyah-Singer index theorem ) for any smooth
gauge background satisfying the topological bound (\ref{eq:topo_bound}).
These constitute the necessary requirements
\cite{twc98:9a} for $ D_c $ to enter (\ref{eq:gen_sol})
such that $ D $ could provide a nonperturbative regularization
for a massless Dirac fermion interacting with a background
gauge field. Explicitly, these necessary requirements are :
\begin{description}
\item[(a)] $ D_c $ is antihermitian
           ( hence $\gamma_5$-hermitian )
           and it agrees with $ \gamma_\mu ( \partial_\mu + i g A_\mu ) $
           in the classical continuum limit.
\item[(b)] $ D_c $ is free of species doubling.
\item[(c)] $ D_c $ is nonlocal.
\item[(d)] $ D_c $ is well defined in topologically trivial background
           gauge field.
\item[(e)] $ D_c $ has zero modes as well as simple poles
           in topologically non-trivial background gauge fields
           ( each zero mode of $ D_c $ must be accompanied by
             a simple pole of $ D_c $ ).
           Furthermore, the zero modes of $ D_c $ satisfy the
           Atiyah-Singer index theorem for any smooth
           gauge background satisfying the bound (\ref{eq:topo_bound}).
\end{description}

The general solution of $ D_c $ satisfying these requirements
had been investigated in ref. \cite{twc99:8}.
However, in general, given any lattice
Dirac operator $ D $, there exists a transformation $ \CT(R_c) $
for $ D $ such that the transformed Dirac operator $ D_c = \CT(R_c) [D] $
is chirally symmetric \cite{twc99:6,twc99:12a}.

\section{ Indices of nonlocal Dirac operators }

In this appendix, we note that the locality condition
{\bf (i)} may not be relevant to the exact zero
modes of a Dirac operator.
Consider a massless fermion in a background gauge field,
the Dirac operator in continuum is
\bea
\label{eq:D_cont_c}
\Dcont = \gamma_\mu ( \partial_\mu + i A_\mu ) \ ,
\eea
which is local, chirally invariant
( $ \Dcont \gamma_5 + \gamma_5 \Dcont = 0 $ )
and antihermitian ( $ \Dcont^{\dagger} = - \Dcont $ ).
Now we can define a new Dirac operator
\bea
\label{eq:D_non}
\Dcont' = \Dcont ( \Id - r b \Dcont )^{-1}
        = ( \Id - r b \Dcont )^{-1} \Dcont \ ,
\eea
where $ r $ is a positive real number and $ b $ denotes a regulator with
mass dimension $ M^{-1} $. One can always choose the value of $ r $ such
that $ \Dcont' $ is nonlocal. Now it is evident that
any zero mode of $ \Dcont $ must be a zero mode of $ \Dcont' $, and
vice versa, i.e.,
\BAN
 \Dcont \phi_{\pm} = 0 &\Leftrightarrow&  \Dcont' \phi_{\pm} = 0,
\EAN
where $ \gamma_5 \phi_{\pm} = \pm \phi_\pm $. Then the number of zero
modes for each chirality must be the same for $ \Dcont $ and $ \Dcont' $,
thus the index of the local Dirac operator $ \Dcont $ is equal to the
index of the nonlocal Dirac operator $ \Dcont' $,
\BAN
 n_{+} ( \Dcont' ) = n_{+} ( \Dcont ), \hspace{4mm}
 n_{-} ( \Dcont' ) = n_{-} ( \Dcont )
\EAN
\beq
\mbox{index}(\Dcont') = n_{-}(\Dcont') - n_{+}(\Dcont') =
 n_{-}(\Dcont) - n_{+}(\Dcont) = \mbox{index}(\Dcont) \ .
\eeq
Therefore, the nonlocal Dirac operator $ \Dcont' $
(\ref{eq:D_non}) has well-defined index, and it also satisfies the
Ginsparg-Wilson relation
\bea
\label{eq:GWR}
\Dcont' \gamma_5 + \gamma_5 \Dcont' = - 2 r b \Dcont' \gamma_5 \Dcont' \ .
\eea
It is interesting to note that the chiral limit ( i.e., $ r = 0 $ )
of the nonlocal $ \Dcont' $ (\ref{eq:D_non}) is the local operator
$ \Dcont $ in (\ref{eq:D_cont_c}),
while on the lattice, the chiral limit of a local
GW Dirac operator $ D $ satisfying {\bf (v)} is
\bea
\label{eq:DcD}
D_c = D ( \Id - a r D )^{-1} = ( \Id - a r D )^{-1} D \ ,
\eea
which must be nonlocal if $ D_c $ is free of species doubling
and has the correct continuum behavior, according to the
Nielson-Ninomiya no-go theorem.  Again, the index of $ D_c $ is
equal to the index of $ D $. Therefore, no matter in continuum or on
a lattice, a nonlocal Dirac operator can have well-defined index,
at least for those obtained by the topologically invariant
transformations, (\ref{eq:D_non}) and (\ref{eq:DcD}).
In other words, if a nonlocal lattice Dirac operator does not have
exact zero modes in topologically nontrivial gauge backgrounds,
then the cause may not be due to its nonlocalness, but its
topological characteristics.

\bigskip
\bigskip
\flushpar
{\bf Acknowledgement }
\bigskip

\noindent
This work was supported by the National Science Council, R.O.C.
under the grant number NSC89-2112-M002-017. I thank Herbert Neuberger
for his helpful comments on the first version of this paper.


\vfill\eject

\end{document}